\documentclass[aps,10pt,pra,twocolumn, groupedaddress,showpacs,showkeys,amsfonts]{revtex4-1}
\usepackage{amsfonts}
\usepackage{amsmath}
\usepackage{amssymb}
\usepackage{graphics,graphicx}
\usepackage{epsf}
\usepackage{subfigure}
\usepackage{hyperref}
\usepackage[all]{hypcap}
\usepackage{color}
\newcommand{\eq}[1]{Eq.~\eqref{#1}}
\newcommand{\eqs}[1]{Eqs.~\eqref{#1}}
\newcommand{\eqss}[2]{Eqs.~\eqref{#1}-\eqref{#2}}
\newcommand{\seq}[1]{Sec.~\ref{#1}}
\newcommand{\app}[1]{App.~\ref{#1}}
\newcommand{\fig}[1]{Fig.~\ref{#1}}

\newcommand{\be}{\begin{equation}}
\newcommand{\ee}{\end{equation}}
\newcommand{\bem}{\begin{multline}}
\newcommand{\bea}{\begin{align}}
\newcommand{\eea}{\end{align}}

\def\cro#1{\left[ #1 \right]}
\def\pare#1{\left( #1 \right)}

\def\abs#1{\lvert{ #1 \rvert}}

\newcommand{\overbar}[1]{\mkern 1.5mu\overline{\mkern-1.5mu#1\mkern-1.5mu}\mkern
1.5mu}

\begin{document}

\title{Tuning nonthermal distributions to thermal ones in time-dependent Paul traps}

\author{H. Landa$^{1}$}
\email{haggaila@gmail.com}
\affiliation{$^1$Institut de Physique Th\'{e}orique, Universit\'{e} Paris-Saclay, CEA, CNRS, 91191 Gif-sur-Yvette, France}

\begin{abstract}

We study the probability distribution of an atomic ion being laser-cooled in a periodically-driven Paul trap using a Floquet approach to the semiclassical photon scattering dynamics. We show that despite the microscopic nonequilibrium forces, a stationary thermal-like exponential distribution can be obtained in the Hamiltonian action, or equivalently in the number of quanta (phonons) of the motion linearized about the  zero of the potential. At the presence of additional stray electric fields, the ion is pushed from the origin of the potential and set into a large-amplitude driven oscillation, and above a threshold amplitude of such ``excess micromotion'', the action distribution of excitations about the driven oscillation broadens and becomes distinctly nonthermal. We find that by a proper choice of the laser detuning the distribution can be made exponential again, with a mean phonon number close to that of the Doppler cooling limit. We derive a relation allowing to deduce just from the experimentally observable photon scattering rate both the required detuning for optimal cooling and the final mean phonon number. These results are important for quantum information processing and other applications, and in particular the derived approach can be applied to crystals of trapped ions in planar configurations, where the driven motion of ions is unavoidable.

\end{abstract}


\maketitle

\section{Introduction and Main Results}\label{Sec:Intro}

A most common method of trapping charged particles in vacuum is by use of electrodynamic Paul traps \cite{paul1990electromagnetic}.  
 The electric potential generated in these traps is time-dependent, as the applied electrode voltages are periodically modulated at a radio-frequency (rf) rate. 
Close enough to the effective potential minimum in many Paul trap variants, the potential is well approximated by a quadrupole term, bilinear in the coordinates. 
A periodically-driven quadrupole potential gives rise to time-dependent, linear, Mathieu equations of motion. The motion of the ions is then characterized by quasi-periodic harmonic oscillations at the so-called ``secular'' frequencies, superimposed with the fast ``micromotion'' driven by the time-dependent potential. Since the equations of motion are integrable (being linear), the description of the motion can be simplified using classical action-angle coordinates. For a given initial condition, the trajectory in the phase-space of coordinates and momenta is restricted to rotations on an invariant manifold. This is a torus whose dimension is half that of the phase-space. The ion's position on the torus at any time is given by the angles which are the generalized coordinates, each evolving independently and increasing linearly in time (with a fixed angular frequency). The measure within the torus (the bounded hyper-volume in the phase-space) is determined by the actions which in the absence of perturbations are conserved quantities of the motion. 

An important tool employed to control the dynamics of an ion and break the conserved quantities is laser cooling, a well established method for removing entropy, e.g.~from a trapped ion's motion \cite{wineland1979laser, javanainen1980,javanainen1980a,javanainen1981laser, gordon1980motion,stenholm1986semiclassical, cirac1994laser, aura2002,leibfried2003,wesenberg2007, epstein2007,marciante2010, PhysRevA.96.012519}. 
 A stochastic process by its very nature, laser cooling is based
on the absorption of photons with a narrow momentum bandwidth determined by the laser, followed by a spontaneous emission of photons with randomly directed momenta. 

In this work we use a recently developed semiclassical framework for studying laser cooling dynamics of ions driven by micromotion up to large amplitudes of motion \cite{rfcooling}. 
We focus on a single ion's distribution in the final stage of the cooling. Since the Paul trap potential is time-dependent, the distribution function in the positions and velocities is non-stationary; it is periodic with the period of the trap \cite{cirac1994laser}. A key idea underlying the current treatment is that by using the action that is a time-independent quantity, the periodic, driven component in the kinetic energy is readily separated from the stochastic component. In terms of the action it is easy to distinguish an effective, stationary thermal-like exponential distribution from other, nonthermal and much more broad distributions possible for the ion in the final stage of Doppler cooling. Such nonthermal distributions arise due to the photon scattering dynamics being modified by a large-amplitude driven motion of the ion, as analyzed in the following.

In order to describe periodically driven photon scattering processes we derive in \seq{Sec:FloquetSolution}, using a Floquet approach, a new solution of the Optical Bloch Equations (OBE) for a two-level system. This Floquet expansion allows us to account accurately for the micromotion, and it can be used quite generally \cite{podlecki2018radiation}. 
In \seq{Sec:Motion} we introduce the setup that is studied in detail in this work, of an ion being laser-cooled in a region of a Paul trap where the potential is approximately a quadrupole potential. 
 We present analytic and numerical results for the simplified case of one-dimensional (1D) motion, with the detailed derivations for three-dimensional (3D) motion left for the Appendix (see below).
In \seq{Sec:Doppler} we review the Fokker-Planck (FP) framework for laser cooling that is the main tool employed in this work. This approach is based on the observation that for typical laser cooling rates we can consider the motion as Hamiltonian over a large number of rotations on the invariant torus, and the laser can be modelled as acting on the action alone (with the angles averaged over). 
The effect of the laser is then to permit the ion to drift and diffuse between the invariant tori of the Hamiltonian phase-space, described by using a Fokker-Planck equation for the probability distribution of the ion in terms of the action.

 \begin{figure}
\includegraphics[width=3.3in]{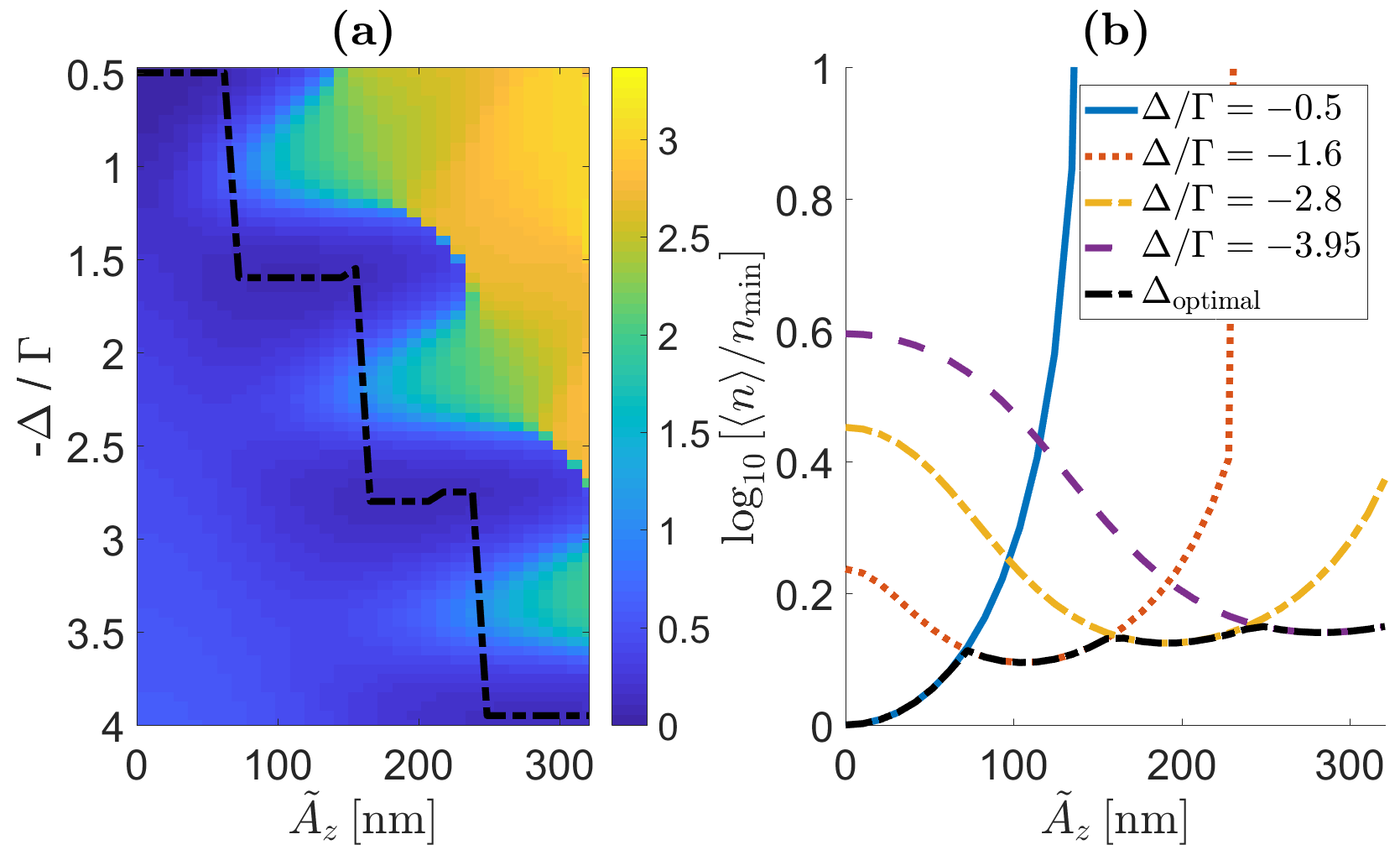}
\caption{(a) The mean steady-state phonon number $\langle n \rangle$ [\eq{Eq:meann}] as a function of the laser detuning $\Delta$ (in units of the natural linewidth of the cooling transition, $\Gamma$), and the excess micromotion amplitude $\tilde{A}_z$ [\eq{Eq:tildez}, given in nanometers for the presented parameters], for one coordinate ($z$) of a $^{24}$Mg$^+$ ion being laser cooled within a linear Paul trap potential with the frequencies given in \eq{Eq:frequencies}. The color code indicates the base-10 logarithm of $\langle n \rangle$, in units of the  Doppler cooling limit phonon number $n_{\min}$, attainable in the upper left corner of the plot, for $\Delta=-\Gamma/2$ and $\tilde{A}_z=0$, and up to $\sim 3$ orders of magnitude larger. The dashed-dotted black line gives the function $\Delta_{\rm optimal}(\tilde{A}_z)$ that determines an optimal detuning as function of $\tilde{A}_z$, for which $\langle n \rangle$ can be reduced to a minimum. (b) Curves obtained from the map of panel (a), showing the attainable minima of the final excitation as a function of the excess micromotion amplitude. Here the vertical axis is truncated at 1, corresponding to $\langle n \rangle=10\times n_{\min}$. }
\label{fig:Imean2}
\end{figure}

 \begin{figure}
\includegraphics[width=3.3in]{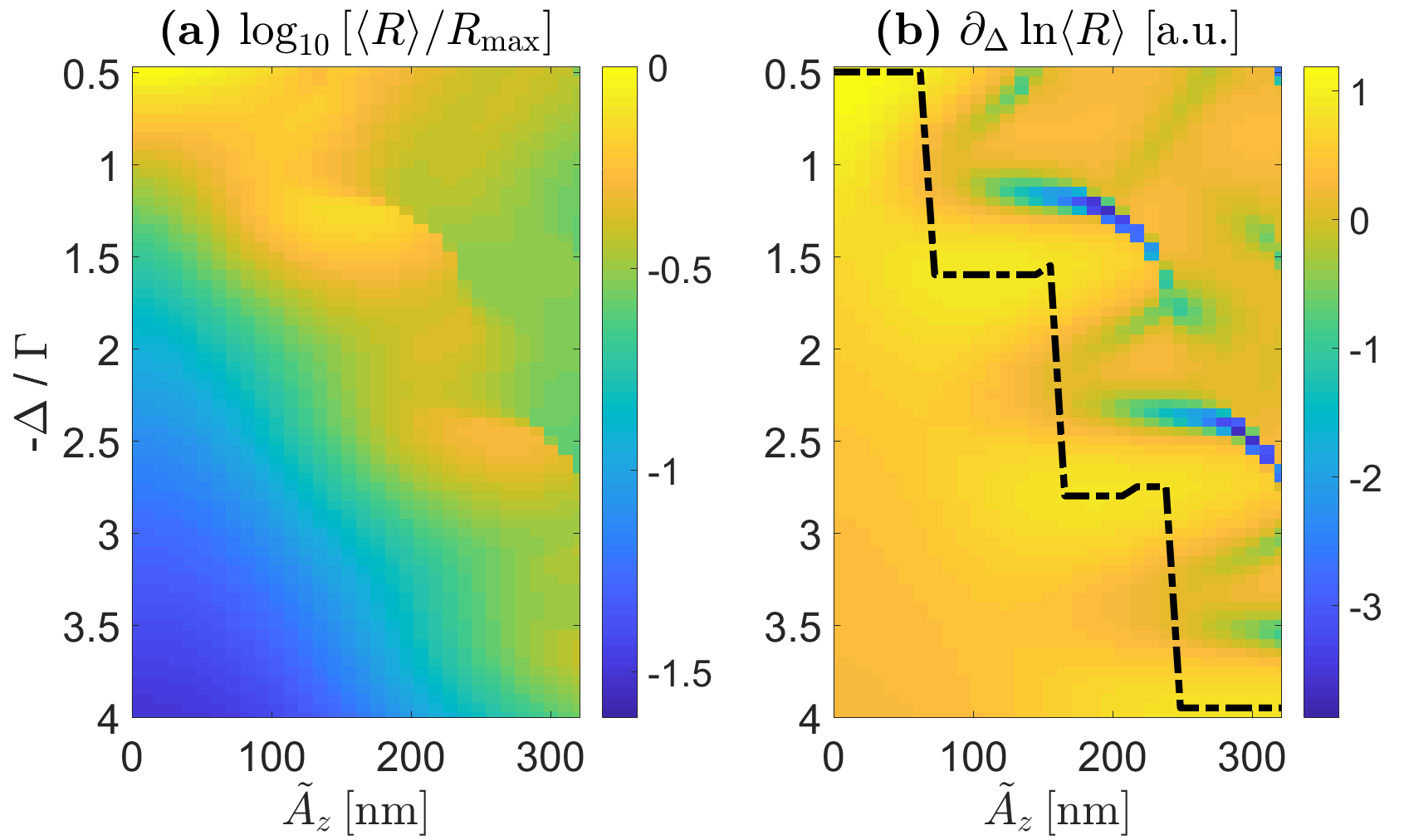}
\caption{(a) The mean photon scattering rate $\langle R \rangle$ [\eq{Eq:meanR}] as a function of the detuning and the excess micromotion amplitude as in \fig{fig:Imean2}. The color code indicates the base-10 logarithm of $\langle R \rangle$, in units of the maximal scattering rate $R_{\max}$, attainable in the upper left corner of the plot, for $\Delta=-\Gamma/2$ and $\tilde{A}_z=0$. (b) The logarithmic derivative (with respect to the detuning) of the mean scattering rate mapped in panel (a). The dashed-dotted black line follows the maximum $\max_\Delta\left[ \partial_\Delta \ln\langle R \rangle\right]$ at each $\tilde{A}_z$ value. This experimental observable gives immediately  both the optimal detuning curve $\Delta_{\rm optimal}(\tilde{A}_z)$ shown in \fig{fig:Imean2}, along which the ion is cooled to a thermal distribution with minimal width. The value of $\langle n \rangle$  along this curve can be obtained as well (see \fig{fig:PI2}). }
\label{fig:Rgrad2}
\end{figure}

The probability distribution in the final stage of Doppler  cooling is known to be thermal for an ion within a time-independent harmonic trap \cite{javanainen1980}. 
As presented in \seq{Sec:CoolingLimit}, we show that even with periodically-driven dynamics where the ion accelerates between the photon absorption and the spontaneous emission, a nearly identical, exponential distribution is obtained with a mean action corresponding to the Doppler limit.
Extending the theory to include stray electric fields which may push the ion away from the origin of the quadrupole potential, leading to ``excess micromotion'' \cite{devoe1989role,Blumel1989,berkeland1998minimization}, detailed numerical calculations are presented in \seq{Sec:Distributions}. Considering motion in 1D along the $z$ coordinate, taken to be a (micromotion-driven) coordinate within a quadrupole Paul trap, we study laser cooling in the low-saturation (low laser intensity) limit. We present results in three different regimes of the micromotion frequency -- low, comparable and high, with respect to the linewidth $\Gamma$ of the electronic transition driven by the laser, approximated as a two-level system.

To summarize the main results of the current study, we consider cooling of a $^{24}{\rm Mg}^+$ ion with trap and laser parameters as detailed in \eqss{Eq:Mathieuparams}{Eq:LaserParams2}. The rf-drive (micromotion) frequency taken here and the oscillator's secular frequency [\eq{Eq:omegaz}] are
\be
 \Omega=2\pi\times 50\,{\rm MHz},\quad
 \omega_z\approx 2\pi\times 2.8\,{\rm MHz}.\label{Eq:frequencies}\ee
 The ion's $z$ motion is expanded in the following form
\be \tilde{z}(t)= \tilde{A}_0+\frac{1}{2}\tilde{A}_z\cos(\Omega t)+\tilde{u}(t),\label{Eq:tildez}\ee
with $\tilde{A}_0$ the ion's mean position, $\tilde{A}_z$ the amplitude of the excess micromotion oscillation, and $\tilde{u}(t)$ the motion expanded about the driven oscillations, which is the free degree of freedom being cooled by the laser.
 We calculate the steady-state of the laser cooled ion when varying two parameters -- the laser detuning $\Delta$ with respect to the resonant rest-frame electronic transition frequency, and the amplitude of the driven excess micromotion oscillations, $\tilde{A}_z$ (reaching 300\,nm for a stray electric field along $z$ of magnitude $\sim 150$V/m). 

Figure \ref{fig:Imean2} shows the mean excitation of the linear oscillator corresponding to the ion's motion, measured in terms of the equivalent mean phonon number $n$ by using the semiclassical relation [see \eq{Eq:quantization}]
\be n \approx I /\hbar,\label{Eq:napprox}\ee where $I $ is the classical action for the linearized motion [$\tilde{u}(t)$], $\hbar$ is Planck's constant, and the zero-point motion [$+1/2$ on the left of \eq{Eq:napprox}] has been neglected. At a fixed value of $\Delta$, a large excess micromotion amplitude could lead to a large phonon number in the steady-state. However, at each fixed excess micromotion amplitude, an optimal value of the detuning (given by $\Delta_{\rm optimal}(\tilde{A}_z)$ and depicted by a dashed-dotted line), can be chosen for which the mean excitation is reduced to a minimal value close to the Doppler cooling limit $n_{\rm min}$ obtained for $\Delta=-\Gamma/2$ and $\tilde{A}_z=0$. 

Noticeable ``tongues'' can be seen in \fig{fig:Imean2}, at intervals obeying $-\Delta=m\Omega$ with $m$ a natural number, i.e.~at the intervals $-\Delta/\Gamma\approx 1.2m$. Since the chosen micromotion frequency is comparable in magnitude to $\Gamma/2$, we account for the photon absorption probability by using a Floquet solution to the periodically-driven OBE equations [\eq{Eq:rhobarsigma}], and the tongues can be attributed to parametric resonances (see also \app{Sec:Bessel}). A brief analysis of the nonmonotonous features of the curves seen in \fig{fig:Imean2}(b) is presented in \seq{Sec:Low}, where we study cooling when approaching the limit of a low micromotion frequency. A further look into the role of the Floquet resonances is presented in \seq{Sec:High}, where the limit of a high micromotion frequency is analyzed.

Figure \ref{fig:Rgrad2}(a) shows the mean photon scattering rate  $\langle R \rangle$ [defined in \eq{Eq:meanR}] in the steady-state calculated in \fig{fig:Imean2}. The importance of this observable which can be straightforwardly measured experimentally is in giving access to both  $\Delta_{\rm optimal}(\tilde{A}_z)$ and $\langle n\rangle$ along this optimal cooling curve. As shown in \seq{Sec:Optimal}, the maximum $\max_\Delta\left[ \partial_\Delta \ln\langle R \rangle\right]$ of the logarithmic derivative (with respect to the detuning) of $\langle R \rangle$ at each fixed $\tilde{A}_z$ value, is obtained along the curve $\Delta_{\rm optimal}(\tilde{A}_z)$ that can hence be readily reconstructed experimentally. In addition, the mean phonon number along this curve can be approximated using the simple formula 
\be \langle n \rangle \approx\left[ \partial_\Delta \ln\langle R \rangle\right]^{-1} \tilde{n},\label{Eq:noptimalgrad}\ee
 where $\tilde{n}$ is a proportionality constant defined in \eq{Eq:ntilde}, which depends only on the trap and laser  parameters.

\begin{figure}
\includegraphics[width=3.2in]{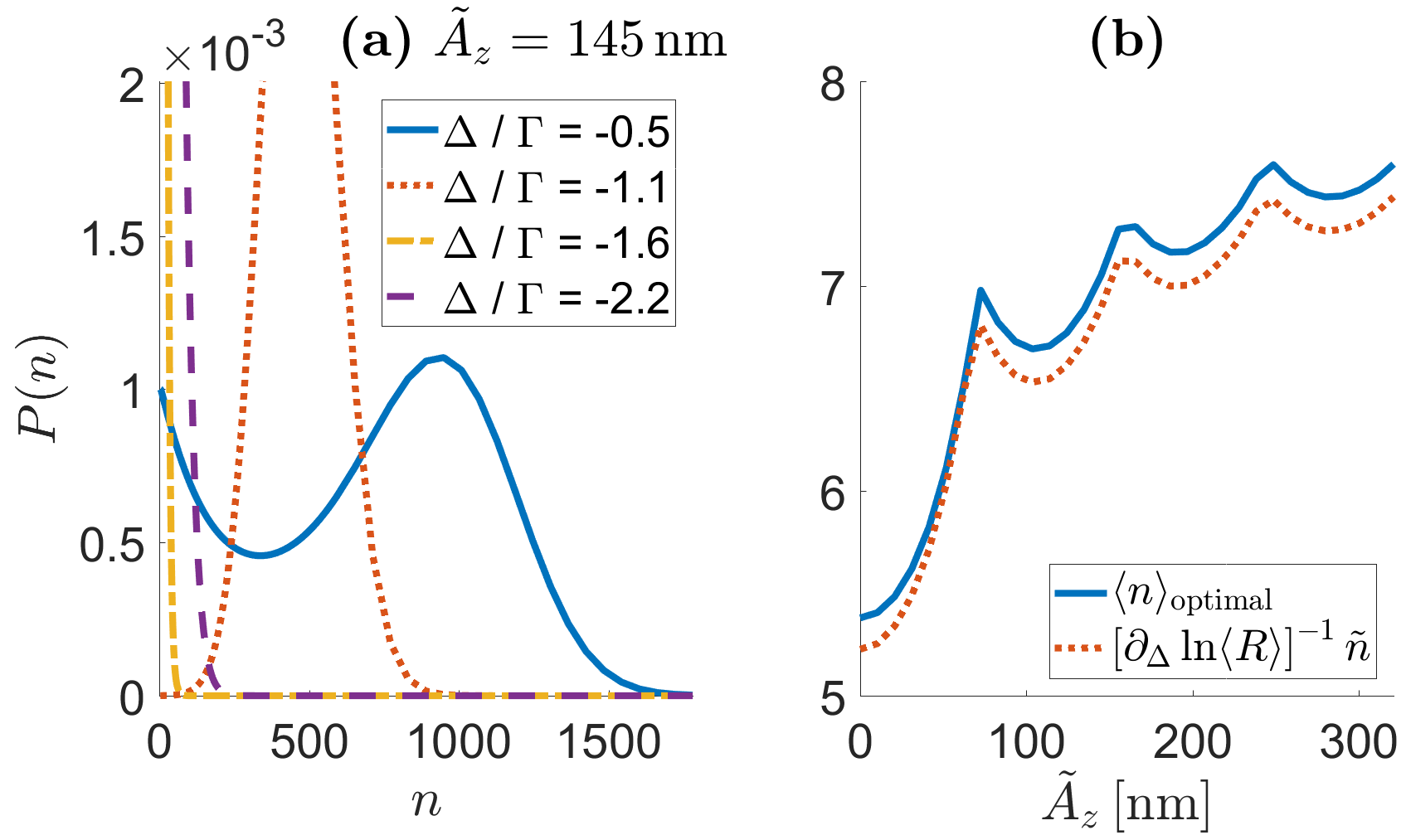}
\caption{(a) The stationary probability distribution  as a function of the phonon number for $\tilde{A}_z=145\,$nm and different values of the detuning $\Delta$, with the laser and trap parameters as in \fig{fig:Imean2}. For $\Delta=-0.5\Gamma$ and $\Delta=-1.1\Gamma$ the distribution is clearly nonthermal. For $\Delta= -1.6\Gamma$ the distribution is exponential (thermal-like) and with the minimal achievable mean phonon number. For higher detunings the distribution broadens (but remains exponential). (b) The optimal mean phonon number as a function of the excess micromotion amplitude [for a detuning chosen along $\Delta_{\rm optimal}(\tilde{A}_z)$], depicted by the solid blue line. The dashed red line gives the optimal mean phonon number as obtained from \eq{Eq:noptimalgrad} which can be directly calculated from the observed photon scattering rate.
}
\label{fig:PI2}
\end{figure}

Figure \ref{fig:PI2}(a) shows a few probability distribution curves for the phonon number with $\tilde{A}_z=145\,{\rm nm}$ and varying values of the detuning. From clearly nonthermal (and even multipeaked) distributions at low detuning, exponential (thermal-like) distributions can be obtained by increasing the detuning. As can be seen in \fig{fig:PI2}(b), along the optimal detuning curve a mean phonon number larger than the minimal value at the Doppler cooling limit by only 1-2 quanta is obtainable for any excess micromotion amplitude. 

Thus we find that $\Delta$ can be used as a control parameter to counteract, to a high extent, the effect of excess micromotion. As shown in \cite{rfcooling}, increasing $\Delta$ is also efficient for cooling the ion from the regime of approximately integrable high amplitude motion in the anharmonic potential of a surface-electrode trap with a high micromotion frequency, wherein a low-detuning laser (with $\Delta\approx -\Gamma/2$) may more easily heat the ion past the trap barrier. We also show in \seq{Sec:High} that such a low-detuning laser may not be efficient in cooling the ion at the presence of stray electric fields and large excess micromotion.
However, it should be noted that a laser beam with a large detuning could actually capture the ion in a large amplitude motion away from the trap centre, when chaotic motion becomes significant in the phase space \cite{rfcycles}.

We conclude with a brief summary and an outlook for possible applications and generalizations in \seq{Sec:Outlook}.
In the Appendix, we lay down the details of the action-angle treatment of a coupled system of Mathieu oscillators. The exact action-angle coordinates (derived in \app{Sec:Mathieu}) can be used for obtaining the FP coefficients for 3D motion of a single ion within quadrupole traps. The derived expressions can be applied to the linearized oscillations of a crystal of trapped ions about their minimum positions \cite{rfions}. Closed form expressions are derived for Doppler cooling in the linear limit (\app{Sec:Final}). Some derivations of the 1D results presented within the main text of the paper are given in \app{Sec:Derivations}, and \app{Sec:Bessel} connects our approach with earlier results in the field \cite{devoe1989role,Blumel1989, berkeland1998minimization}.


\section{The Optical Bloch Equations}\label{Sec:OBE}

\subsection{Derivation}\label{Sec:Interaction}

In this subsection we review the derivation of the Optical Bloch Equations (OBE) following \cite{grynberg2010introduction}.
A monochromatic laser in a travelling-wave configuration can be described by a classical electric field of the form
\be \vec{E}_{\rm L}\left(\vec{r},t\right)=E_{\rm L}\hat{e}\cos\left(\omega_{\rm L}t -\vec{k}\cdot\vec{r}+\phi_{\rm L}\right),\ee
where $E_{\rm L}$ is the electric-field amplitude, $\hat{e}$ the unit-vector of its polarization, $\vec{k}$ its wavevector, $\omega_{\rm L}=|\vec{k}|/c$ its frequency, $\phi_{\rm L}$ the optical phase, and $\vec{r}$ the ion's position vector.
The OBE for a two-level system describe the time evolution of the elements of the density matrix of the two-level system approximating the electron, subject to the driving by the laser and to spontaneous emission. The electronic levels are the ground-state $|g\rangle$ with zero energy and the excited level $|e\rangle$, with an electronic transition between the two of frequency $\omega_e$. The OBE are derived in the long-wavelength approximation, such that the laser wavelength is much larger than the atomic size.
The two-level Hamiltonian is
\be H_e = \hbar \omega_e\left|e\right\rangle\left\langle e\right|+\hbar\Omega_{\rm R}\cos\left(\omega_{\rm L}t+\phi\right)\left[\left|g\right\rangle\left\langle e\right|+ \left|e\right\rangle\left\langle g\right| \right],\label{Eq:He}\ee
where $\Omega_{\rm R}$ is the Rabi frequency (proportional to $E_{\rm L}$ and to the matrix element of the electric dipole transition), and 
\be \phi=\phi(\vec{r})=-\vec{k}\cdot\vec{r}(t)+\phi_{\rm L},\label{Eq:phi}\ee depends on the ion's position in configuration space.

 The density-matrix of the two-level system evolves subject to the Hamiltonian of \eq{Eq:He}, and with spontaneous emission described by Lindblad-type terms. The master equation, in the rotating-wave approximation, can be written in terms of the four matrix elements  $\sigma_{ab}$, with $a,b\in \{g,e\}$, which gives the OBE in the form 
\begin{align}
&\dot{\sigma}_{ee} =-\Gamma{\sigma}_{ee} -i\frac{\Omega_{\rm R}}{2} \left(e^{-i\omega_{\rm L}t-i\phi}{\sigma}_{ge} - e^{i\omega_{\rm L}t+i\phi}{\sigma}_{eg}\right),\label{Eq:OBE1}\\
&\dot{\sigma}_{ge} = \left(i\omega_{e}-\frac{\Gamma}{2}\right){\sigma}_{ge} -i\frac{\Omega_{\rm R}}{2} e^{i\omega_{\rm L}t+i\phi} \left({\sigma}_{ee}-{\sigma}_{gg}\right),\label{Eq:OBE2}
\end{align}
where ${\sigma}_{ge}={\sigma}_{eg}^*$ and from the density matrix having unit trace we get ${\sigma}_{gg}+{\sigma}_{ee}=1$, which implies ${\sigma}_{ee}-{\sigma}_{gg}=2{\sigma}_{ee}-1$. Hence \eqss{Eq:OBE1}{Eq:OBE2} form a closed system, that depends on the (semiclassical) dynamics of the ion through the variation of the phase of \eq{Eq:phi}. It can be seen that the off-diagonal terms ${\sigma}_{ge}$ and ${\sigma}_{eg}$ decay at a rate $\Gamma/2$, while the diagonal terms decay with rate  $\Gamma$, where ${\Gamma}$ is the spontaneous emission rate (the linewidth of the transition).

Substituting into \eq{Eq:OBE2} the new variable
\be\sigma_{ge}'= e^{-i\omega_{\rm L}t -i\phi\left[\vec{r}(t)\right]}\sigma_{ge},\label{Eq:OBEsubs1}\ee
we get after rearranging, the modified OBE
\begin{align}
&\dot{\sigma}_{ee} =-\Gamma{\sigma}_{ee} -i\frac{\Omega_{\rm R}}{2} \left({\sigma}_{ge}'- {\sigma}_{eg}'\right),\label{Eq:OBE1b}\\
&\dot{\sigma}_{ge}' = \left[-i\left(\Delta-\vec{k}\cdot\vec{v}\right)-\frac{\Gamma}{2}\right]{\sigma}_{ge}' -i\frac{\Omega_{\rm R}}{2} \left(2{\sigma}_{ee}-1\right),\label{Eq:OBE2b}
\end{align}
with the detuning $\Delta=\omega_{\rm L}-\omega_e$ and the velocity $\vec{v}=\dot{\vec{r}}$.

When $\vec{v}$ can be assumed to evolve slowly on the scale of $\Gamma/2$, the steady-state of the coupled OBE can be found by setting the time-derivatives to 0, and ${\sigma}_{ee}$, which is the mean probability of the electron to be found in the excited level, obtains a well-known Lorentzian form in the velocity \cite{rfcooling} with the saturation accounted for).

\subsection{Low-saturation solution with a Floquet approach}\label{Sec:FloquetSolution}

To proceed we now consider the low saturation limit (low laser intensity, or a solution to leading (second) order in $\Omega_{\rm R}$), 
\be s=2\left(\abs{{\Omega}_{\rm{R}}}/{\Gamma}\right)^2\ll 1.\label{Eq:sll1}\ee
 In this limit, we assume that the solution obeys $\sigma_{ee}\ll 1$ and write in \eq{Eq:OBE2b}, $2{\sigma}_{ee}-1\approx -1$.
Then, \eq{Eq:OBE2b} decouples to become
\be \dot{\sigma}_{ge}' = \left[-i \Delta_{\rm eff}(t)-\frac{\Gamma}{2}\right]{\sigma}_{ge}' +i\frac{\Omega_{\rm R}}{2},\label{Eq:OBE2c}\ee
where we have defined 
\be \Delta_{\rm eff}(t)=\Delta-\vec{k}\cdot\vec{v}(t).\label{Eq:Deltaeff_t}\ee
The general solution of \eq{Eq:OBE2c} consists of the sum of a transient solution to the homogeneous part of the equation which will decay with an exponential envelope of $e^{-\Gamma t/2}$, and a particular solution to the inhomogeneous \eq{Eq:OBE2c} which gives the steady-state. The solutions can be obtained in closed-form by integration, however, it is instructive to use a Floquet approach to analyze the steady-state solution.

Let us assume that the velocity (along the laser wavevector) is periodic in time with $\Omega$ the fundamental frequency of the periodic motion.
Using the nondimensional units given in \app{Sec:Mathieu}, we set $\Omega=2$ obtained by rescaling 
\be t\to\Omega t/2\label{Eq:OmegaRescale}\ee and measuring accordingly all frequency quantities in units of $\Omega/2$,
and hence we can write
\be -i\Delta_{\rm eff}(t) -\frac{\Gamma}{2}= \sum_{n}D_{2n}e^{i2n t}.\label{Eq:DeltaEffOmega}\ee
The particular non-transient solution to the inhomogeneous \eq{Eq:OBE2c} which is periodic with the parametric drive can be written using a Floquet expansion as
\be {\sigma}_{ge}'=\sum_{n}G_{2n}e^{i2n t},\ee
with coefficients $G_{2n}$ that will be determined in the following.
Substitution in \eq{Eq:OBE1b} gives
\be \dot{\sigma}_{ee} =-\Gamma{\sigma}_{ee} +\Omega_{\rm R}{\rm Im} \left(\sum_{n}G_{2n}e^{i2n t}\right),\ee
whose particular periodic solution takes the form
\bem
{\sigma}_{ee}=\Omega_{\rm R}e^{-\Gamma t}\sum_{n} \int^t {\rm Im} \left(G_{2n} e^{(\Gamma +i2n) t'}\right) dt'=\\
=\sum_{n}\frac{\Omega_{\rm R}}{\Gamma^2 +(2n)^2}{\rm Im} \left( G_{n} \left(\Gamma -i2n\right) e^{i2n t}\right).\label{Eq:sigmaFloquet}\end{multline}

The Floquet expansion coefficients $G_{2n}$ can be found using known methods. In the important case of a simple harmonic and time-reversal invariant $\Delta_{\rm eff}$, we can use known expressions for an inhomogeneous driven Mathieu oscillator \cite{rfmodes}. We assume the following expansion for \eq{Eq:DeltaEffOmega},
\be -i\Delta_{\rm eff}(t) -\frac{\Gamma}{2}= D_0-2D_2 \cos(2t),\label{Eq:D_nExpansion}\ee
and substitute in \eq{Eq:OBE2c} to obtain 
\bem 
-i\sum_{n}2n G_{2n}e^{i2n t}+\\ \left[D_0-D_2 (e^{i2t}+e^{-i2t})\right]\sum_{n} G_{2n}e^{i2n t} = -i\frac{\Omega_{\rm R}}{2},\end{multline}
which implies the recursion relations
\be -i2nG_{2n}+D_0 G_{2n} -D_2\left(G_{2n-2}+G_{2n+2}\right)=-i\frac{\Omega_{\rm R}}{2}\delta_{n,0}.\ee
Defining
\be R_{2n}=D_0-i2n,\ee
the equations for $n\ge 1$ give the continued fraction
\be G_{2}=T_2 D_2 G_0,\ee
with 
\be T_2=[R_2-D_2[R_4-D_2[R_6-...]^{-1} D_2]^{-1} D_2]^{-1}.\ee
The equations for $n\le -1$ can be rearranged to get a similar relation that progresses towards negative $n$ values,
\be D_2 G_{2n+2} = R_{2n} G_{2n} -D_2 G_{2n-2} ,\qquad n\le -1,\label{Eq:Recur2}\ee
which gives the continued fraction
\be G_{-2}=T_{-2} D_2 G_0,\ee
with 
\be T_{-2}=[R_{-2}-D_2[R_{-4}-D_2[R_{-6}-...]^{-1} D_2]^{-1} D_2]^{-1},\ee
and the solution of the recursion relations is
\be G_0=-i\frac{\Omega_{\rm R}}{2}\left[D_0-D_2\left( T_2+T_{-2}\right) D_2\right]^{-1},\label{Eq:G0solution}\ee
from which the rest of the coefficients follow immediately using, e.g., \eq{Eq:Recur2}.

For the excited level excitation probability we get 
the particular periodic solution taking the form
\be
{\sigma}_{ee}(t)
=\sum_{n}\frac{\Omega_{\rm R}}{\Gamma^2 +(2n)^2}{\rm Im} \left[ G_{2n} \left(\Gamma -i2n\right) e^{i2n t}\right].\label{Eq:sigmaeet}\ee
When only the average over the period of the driven motion is relevant, it is 
the single $n=0$ Floquet component that determines the result 
\be \bar{\sigma}_{ee}=
\frac{1}{T}\int_0^T\sigma_{ee}(t)dt
=\frac{\Omega_{\rm R}}{\Gamma}{\rm Im} G_{0}\label{Eq:sigma0bar},\ee
with the rescaling of \eq{Eq:OmegaRescale} implying that in the units used above,
\be T=2\pi/\Omega=\pi.\ee
In general, $G_0$ would still be given implicitly, requiring to solve the recursion relations, using \eq{Eq:G0solution}.
In the limit that the periodically-driven motion can be considered as frozen during the time-scale of the internal level dynamics, i.e.~for
\be \Omega/2\pi\ll \Gamma/2\label{Eq:LowOmega},\ee
the solution reduces to the known (low-saturation) Lorentzian probability of absorption of a photon \cite{rfcooling}, which can be obtained by setting $D_2\to 0$ above,
\be
{\sigma}_{ee}^0
=\frac{\Omega_{\rm R}^2}{\Gamma^2 +(2\Delta_{\rm eff})^2} = \frac{s/2}{1+\pare{2\Delta_{\rm{eff}}/\Gamma}^2}.\label{Eq:rho_p_z}\ee
Even if the condition in \eq{Eq:LowOmega} does not hold, in the limit of small amplitude of the motion, the limit $D_2\to 0$ can be taken and the expression in \eq{Eq:rho_p_z} can be linearized in the velocity and used, as discussed below.

\section{Motion in a quadrupole trap with excess micromotion}\label{Sec:Motion}

We start with the $z$ motion of an ion in a 1D time-dependent quadrupole potential, displaced by a constant electric field $E_z$,
 \be {V}_{\rm{e}}= \frac{1}{2}(a_z-2q_z\cos2t) z^2-E_z z.\label{Eq:Mo}\ee
The units are nondimensional, obtained by rescaling the time as in \eq{Eq:OmegaRescale} by half the trap's driving frequency, $\Omega/2$, and rescaling the coordinate by a natural lengthscale $w$ (discussed below), absorbing also the ion charge $e$ and mass $m$ into the nondimensional parameters, $a_z$, $q_z$, and $E_z$ in a standard way \cite{leibfried2003,rfions,rfchaos,rfcooling}, detailed  also in \app{Sec:Mathieu}.
The solution of the inhomogeneous linear Mathieu equation of motion (e.o.m) derived from ${V}_{\rm{e}}$ can be written as \cite{rfmodes}
\be z(t)=\bar{z}(t)+u(t),\qquad \bar{z}=\sum_n B_{2n}e^{i2nt},\label{eq:zt}\ee
 where $n\in\mathbb{Z}$ and $B_{2n}=B_{-2n}$.
 The term $\bar{z}(t)$, being $\pi$-periodic (since the rf drive frequency is 2 in rescaled units) is known as excess micromotion as it can (typically) be minimized, by using controlled electric fields (that make $E_z$ effectively small). Although it cannot be cooled away [in contrast to $u(t)$], $\bar{z}(t)$ is completely coherent and not stochastic \cite{devoe1989role, 
zigzagexperiment}, and in this work we assume that it is invariant under time-reversal.
Equation \eqref{eq:zt} defines a time-dependent canonical transformation to the new coordinate $u(t)$, with the conjugate momentum 
\be p=\dot{u},\ee (with the mass equal to 1). 
The transformed Hamiltonian becomes
\be H_0 ({u},{p},t)=\frac{1}{2}{p}^2+V_{\rm M.o.}({u},t),\label{Eq:H0_1D}\ee
with the nondimensional Mathieu oscillator potential
\be V_{\rm M.o.}(u,t) = \frac{1}{2}(a_z-2q_z\cos2t) u^2.\ee

 We can now introduce the action-angle variables $I$ and $\theta$, defined by a second (time-dependent) canonical transformation, using two functions of the phase space and time, 
\be I=\Lambda({u},{p},t),\qquad \theta=\Theta ({u},{p},t).\label{Eq:Itheta} \ee
The action-angle variables constitute a very useful choice of variables, since $I$ is conserved during the Hamiltonian motion in the time-dependent potential, in contrast to the energy. 
The exact transformation functions are given in \app{Sec:Hamiltonian1D}. To the leading order in $q_z$ we can write
\bea
u&\approx \sqrt{{2I}/{\nu_z}}\cos\theta, \\ p&\approx -\sqrt{2I\nu_z}\sin\theta + q_z\sin(2t) \sqrt{{2I}/{\nu_z}}\cos\theta,\label{Eq:transf}\end{align}
with 
\be\theta(t)=\nu_z t +\phi,\label{Eq:thetadef1D}\ee
 where $\phi$ is determined by the initial conditions, and the characteristic exponent $\nu_z(a_z,q_z)$ of the Mathieu equation can be approximated by
\be \nu_z\approx\sqrt{a_z + q_z^2/2}, \qquad a_z,q_z^2 \ll 1,\label{Eq:nuapprox}\ee
 which determines the secular oscillation frequency in physical units, 
 \be \omega_z=\nu_z\Omega/2\label{Eq:omegaz}.\ee
  To the same accuracy we have for the coefficients of $\bar{z}$ of \eq{eq:zt}, 
\be B_0 \approx  E_z/\nu_z,\qquad B_2\approx -B_0 q_z /4,\label{Eq:B0B2}\ee
 with the rest of the series truncated (\app{Sec:Hamiltonian1D}). 
 We define the amplitude of excess micromotion to be the peak-to-peak oscillation due to $E_z$, given to this order by
  \be A_z=4B_2= 2q_z E_z/\nu_z,\label{Eq:Az}\ee
  which is linear in $E_z$.
At the same order we can approximate
\be \dot{\bar{z}}(t) =-\sum_{n>0}4nB_{2n}\sin(2nt)  \approx -A_z \sin(2t).\label{Eq:dotbarzapprox}\ee

\section{Laser cooling}

\subsection{Laser cooling in the finite lifetime  treatment}\label{Sec:Doppler}

In this subsection we review the semiclassical laser cooling framework developed in \cite{rfcooling}, as it is employed in this work in combination with the Floquet solution of the OBE presented in \seq{Sec:FloquetSolution}, and applied to the motion in a quadrupole trap as described in \seq{Sec:Motion}. This approach is based on conservation of energy and momentum at each photon absorption event and each spontaneous emission occurring after a random delay due to the nonzero lifetime of the electronic excited level.
 Between the absorption and emission, we assume that the ion moves completely classically on an invariant torus of the Hamiltonian phase-space, and is decoupled from the electromagnetic field. 
The presented theory is valid in the limit of a low saturation of the transition [\eq{Eq:sll1}], when the ion spends most of the time in its electronic ground-state, which is often chosen in practice for allowing to reach the lowest cooling limit.

 For the 1D treatment presented in following, when the stochastic dynamics are slow in comparison with the Hamiltonian motion, integrating over the angle $\theta$ allows one to obtain an effective Fokker-Planck equation for the probability  distribution $P(I,t)$ 
\begin{align}
\frac{\partial {P}(I,t) }{\partial t} = -\frac{\partial {S}({I},t) }{\partial I}\equiv - \frac{\partial }{\partial I} \cro{\Pi_I {P}} +\frac{1}{2} \frac{\partial^2 }{{\partial I}^2} \cro{\Pi_{II}{P}},
\label{eq:FP2}
\end{align}
with $S$ the probability flux.
Denoting with an overbar the torus average over any function $\Xi[{I},{\theta},t]$ of the phase space which is assumed to have some arbitrary period $T$;
\be \Xi\cro{{I},{\theta},t+T}=\Xi\cro{{I},{\theta},t},\ee 
we define
\be
\overline{\Xi}(I)\equiv \frac{1}{T} \int_0^{T}dt\frac{1}{2\pi }\int \Xi\cro{{I},{\theta},t}d{\theta}.\label{Eq:TorusAverageDef1D}
\ee  
The action drift and diffusion coefficients, respectively, entering \eq{eq:FP2} for the laser cooling setup considered here are
\bea &\Pi_{I}(I) =\overline{\Gamma\rho\left[
p_{\rm r} \frac{\partial \Lambda}{\partial p} +\frac{1}{2} p_{\rm r}^2   \left(   \frac{\partial^2 \Lambda}{\partial p^2}+\mu \left\langle\frac{\partial^2 \Lambda}{\partial p^2} \right\rangle_\Gamma  \right) \right]},\label{Eq:PiIc1D}\\
&\Pi_{II}(I) =  \overline{\Gamma\rho  \left[p_r^2 \left(\frac{\partial\Lambda}{ \partial p}\right)^2 + p_r^2 \mu  \left\langle\left(\frac{\partial\Lambda}{ \partial p}\right)^2 \right\rangle_\Gamma 
\right] },
\label{Eq:PiIIc1D}\end{align}
where $\mu$ is a constant of order unity \cite{rfcooling}, which stems from the second moment of the dipole radiation pattern (and is often denoted by $\alpha$ \cite{javanainen1981laser} or $\xi$ \cite{leibfried2003} in the existing literature), and ${p}_{\rm{r}}={\hbar} {k}$ is the photon recoil momentum with ${\hbar}$ rescaled as in \eq{Eq:hbar}.
The other terms in the right hand side of \eqss{Eq:PiIc1D}{Eq:PiIIc1D} are all functions of the phase space time and point where a photon absorption occurred, 
 \be Z_a\equiv\{u,p,t\},\label{Eq:Za}\ee
that is averaged on a given torus by the definition  in \eq{Eq:TorusAverageDef1D}.
The required derivatives of the action-angle transformation function $\Lambda(u,p,t)$ [defined in \eq{Eq:Itheta}], can be found in \app{Sec:Mathieu}, and to the leading order in $q_z$ [\eq{Eq:nuapprox}], coincide with  the simple formulae for the harmonic oscillator,
\be  \frac{\partial \Lambda}{\partial p} \approx {-\sqrt{2I/\nu_z}\sin\theta},\qquad   \frac{\partial^2 \Lambda}{\partial p^2} \approx \frac{1}{\nu_z}.\label{Eq:LambdaMaDerivatives}\ee
We stress that the approximation in \eq{Eq:LambdaMaDerivatives} gives accurate results  for the excess micromotion (with a neglected contribution approximately equal to $q_z^2/4$).

The rate of photon absorption-emission cycles in the low-saturation limit as a function of the phase-space point is given by $\Gamma\rho$, where in general one should take $\rho =\sigma_{ee}(t)$, the excited level population defined in \seq{Sec:OBE}. The different limits of $\sigma_{ee}(t)$ employed in this work are discussed in \seq{Sec:Rho}.
The emission phase space point is averaged over through the integration of the waiting time distribution $\langle\cdot\rangle_\Gamma$. Given an absorption that occurred at the phase-space point $Z_a$,
  the mean value of any function of the phase space, at the time of emission, is given by
\be\left \langle \Xi(Z_a) \right\rangle_{\Gamma} \equiv  \int_0^\infty \Gamma e^{-\Gamma t'} \Xi(Z(t+t'; Z(t)=Z_a))dt'.\label{Eq:GammaDecay}\ee
The time integral in \eq{Eq:GammaDecay} is to be performed along the trajectory, denoted with the notation of \eq{Eq:Za} as $Z(t+t')$, as it starts at the phase space point $Z_a$ and evolves according to the Hamiltonian motion at fixed ${I}$. The zero lifetime limit can be obtained from the treatment above if in \eq{Eq:GammaDecay} it can be assumed \cite{rfcooling} that $\Gamma e^{-\Gamma t'}\approx \delta(t')$. In addition, an adiabaticity condition is assumed that justifies the averaging of \eq{Eq:TorusAverageDef1D}. A simple criterion is obtained by requiring a small relative change in action due to both
drift and diffusion, during a cycle of the motion:
\be \Pi_{I}/\nu_z\ll I,\qquad \Pi_{II}/\nu_z\ll I^2.\ee

A steady-state of the FP equation with appropriate boundary conditions can be obtained by setting the left-hand-side to 0, and then noting that the reflecting boundary condition at the origin [$S(I=0,t)=0$] implies that this would be a zero-current state [$S(I,t)=0$]. 
Integrating the resulting equation gives the time-independent distribution
\be P(I)=\mathcal{N}\left[\Pi_{II}(I)\right]^{-1}\exp\left\{2\int^I \frac{\Pi_I(I')}{\Pi_{II}(I')} dI'\right\},\label{Eq:PI}\ee
with $\mathcal{N}$ the normalization factor. This steady-state solution is relevant if the exponential decays fast enough (as a function of $I$), implying physically that indeed the ion remains trapped for a long time in a bounded region of phase space.

\subsection{The photon absorption probability}\label{Sec:Rho}

Using the steady-state distribution $P(I)$ of \eq{Eq:PI} we can define the mean value of any function of the phase-space (averaged over the angles), e.g.~the mean action
\be \langle I\rangle\equiv\int_0^\infty I P(I)dI\label{Eq:meanI},\ee
which, using \eq{Eq:napprox} that neglects the zero-point motion, gives immediately the mean phonon number
\be \langle n\rangle\equiv\int_0^\infty I P(I)dI/\hbar \label{Eq:meann}.\ee
The mean photon scattering rate is also important, and in particular it can be measured experimentally,
\be \langle R\rangle \equiv \int_0^\infty \overline{\Gamma\rho} P(I)dI.\label{Eq:meanR}\ee

Finally, we are left with the need to specify the photon absorption probability function $\rho$. For the 1D dynamics studied here, 
the effective detuning due to the Doppler shift is
\be \Delta_{\rm{eff}}=\Delta-{k}{v_z}({p},t)\label{Eq:Deltaeff1D},\ee
with ${k}$ being the 1D laser wavenumber, and the real-space velocity ${v_z}({p},t)$ that is a function of the canonical momentum ${p}(t)$ expanded about $\bar{z}(t)$ and the time $t$, is given [using \eq{eq:zt}] by 
\be {v_z}=\dot{z}(t) =\dot{\bar{z}}(t)+p(t).\label{Eq:vzexpand}\ee
To leading order in $q_z$, using \eqss{Eq:transf}{Eq:thetadef1D}  and \eq{Eq:dotbarzapprox}, and under the assupmtion that $\nu_z/2\pi\ll \Gamma/2$ which allows to treat $\theta$ as frozen on the time-scale leading to a stationary state in the internal dynamics, we see that \eq{Eq:Deltaeff1D} can be plugged in \eq{Eq:D_nExpansion} with the expansion coefficients
\bea D_0 =-i(\Delta+k\sqrt{2I\nu_z}\sin\theta)-\Gamma/2 ,\\ D_2=-ik(A_z-q_z \sqrt{2I/\nu_z}\cos\theta)/ 2.\label{Eq:D2} \end{align}
Thus, for the most general case (in the low-saturation limit) we should plug  $\rho=\sigma_{ee}(t)$ of \eq{Eq:sigmaeet} in \eqss{Eq:PiIc1D}{Eq:PiIIc1D}. With the approximation of \eq{Eq:LambdaMaDerivatives}, none of the other terms (except $\rho$) in \eqss{Eq:PiIc1D}{Eq:PiIIc1D} depend on $t$, and hence the average over a micromotion period in \eq{Eq:TorusAverageDef1D} can be carried out independently as in \eq{Eq:sigma0bar}, resulting in the simpler expression 
\be \rho=\bar{\sigma}_{ee}(p).\label{Eq:rhobarsigma}\ee
In the limit $D_2\to 0$, we can use \eq{Eq:rho_p_z};
\be\rho\to {\sigma}_{ee}^0(v_z).\label{Eq:rhotosigma0}\ee
This limit can be taken when both the excess micromotion amplitude and the action are low enough (the cooling limit without excess micromotion, treated analytically in \seq{Sec:CoolingLimit}), and when the micromotion frequency is low enough, when the condition of \eq{Eq:LowOmega} holds.

\subsection{The cooling limit without excess micromotion}\label{Sec:CoolingLimit}

In this subsection we summarize the results of a derivation of the Doppelr cooling limit presented in detail for the general 3D motion as in \app{Sec:Final}. In the absence of excess micromotion, the ion velocities can be assumed to be small near the origin of the quadrupole trap. This defines a specific limit of the cooling, that can in fact be analyzed analytically. 
In this limit the Lorentzian $\rho={\sigma}_{ee}^0$
of \eq{Eq:rho_p_z} can be used as discussed above, and can be linearized in the velocity, subject to the condition
\be I \ll I_{\rm linear} = \left(\frac{\Gamma^2+4\Delta^2}{8k\Delta}\right)^2\frac{1}{2\nu_z}.\label{Eq:Ilinear}\ee
In this approximation, we can write for the action drift and diffusion coefficients (with $s\ll 1$), using the expressions given in \app{Sec:Final} and using \eq{Eq:UtVsIdentity},
\be \Pi_{I}^l = \gamma I+\frac{1}{2}i_z,\quad \Pi_{II}^l =i_z I ,\label{Eq:PiIl1D}\ee
with
\begin{align}
i_z= p_{\rm r}F_{\rm r}(1+\mu) c_z, \label{Eq:FP1Dh}
\end{align}
where the mean radiation force and the rate of momentum damping having the well-known forms (\cite{javanainen1980,leibfried2003,rfcooling}, respectively,
\be F_{\rm r}=\frac{p_{\rm r}\Gamma s/2}{1 +(2\Delta/\Gamma)^2}, \quad \gamma = \frac{4k p_{\rm r}s \Delta/\Gamma}{\left[1 +(2\Delta/\Gamma)^2\right]^2},\label{Eq:F0gamma1D}\ee
$i_z$ generalizes a similar coefficient for the harmonic oscillator \cite{rfcooling},
and $c_z$ is defined in \eq{Eq:c_z}, while to the leading order in $q_z$, using \eqs{Eq:nuapprox} and \eqref{Eq:LambdaMaDerivatives}, we have
\be c_z\approx \nu_z^{-1}.\ee

Then for $\Delta<0$ the final stage of the cooling reduces to an exponential, thermal equilibrium-like distribution in the conserved action, that is nearly identical to the well-known Doppler cooling limit that is obtained in limit of a vanishing excited level lifetime within a static harmonic potential (the heavy particle \cite{javanainen1980}). The equilibrium is
determined by the balance of momentum dissipation and diffusive heating. It is a zero-current distribution in the action, given by
\be P(I)=\lambda_z e^{-\lambda_z I},\ee
where $\lambda_z=-2\gamma/i_z$, which gives the mean and standard deviation of the action, $\langle I\rangle =\sqrt{\langle (I-\langle I\rangle)^2\rangle }= \lambda_z^{-1}$, can be obtained from \eq{Eq:Imeanlambda} and reads here
\be I_{\rm limit}\equiv \lambda_z^{-1}= \hbar \frac{\Gamma}{8}c_z ( 1 + \mu) \left[ \frac{\Gamma}{2 \left|\Delta\right|} + \frac{2 \left|\Delta\right|}{\Gamma}\right].\label{Eq:Ilimit} \ee

Although the strandard Doppler cooling limit \cite{leibfried2003} is derived by assuming an ion in a time-independent potential and being stationary in space during a photon absorption-emission cycle, the final distribution derived here, for motion with micromotion, turns out to be thermal as well, and coincides with that of the harmonic oscillator of the same secular frequency, up to the correction due to $c_z$ not being exactly $1/\nu_z$. These results have been found numerically in the example studied in \cite{rfcooling}, and here it is derived for the general case.
 The fact that the ion can emit the photon at a random time along the trajectory after the absorption, and hence with the ion's a momentum (or kinetic energy) being very different than that at the absorption, does not lead on average to an increase in the action diffusion due to the scattering events. 
 The analysis in \app{Sec:Final} indicates that this is a consequence of the memoryless decay process together with the linearity of the oscillations, and the linearity of the Lorentzian (in the small velocity limit). Under such conditions, the phase-space averaging of such scattering events (with the accessible ion momenta), cancels out the effects of micromotion (to leading order).
  It serves to show the strength of the action-angle as the right choice of coordinates for analyzing the motion in the time-dependent potential. 

\subsection{Optimal cooling with excess micromotion}\label{Sec:Optimal}

In this subsection we consider the optimal choice of detuning for cooling the ion to the narrowest (exponential) distribution in action, in the 1D setting. Assuming that an exponential distribution can indeed be obtained (as shown using the numerical calculations presented in \seq{Sec:Intro} and \seq{Sec:Distributions}), it can be characterized as just by using the observable photon scattering rate. The following derivation does not assume a specific form for the photon absorption probability, only that it is a function of the Doppler shift defined in \eq{Eq:Deltaeff_t}, i.e.
\be \rho=\rho(\Delta_{\rm eff})=\rho( \Delta - k\dot{\bar{z}}-kp).\ee
In the following we will refer to the two partial derivatives
\be \partial_\Delta \rho=\rho'(\Delta_{\rm eff}),\quad  \partial_p \rho=-k\rho'(\Delta_{\rm eff}),\ee
where $\rho'$ denotes the derivative with respect to the argument. Defining the Doppler shift at $p=0$, due only to excess micromotion,
\be \Delta_{\rm e}(t)=\Delta - k \dot{\bar{z}}(t),\label{Eq:Deltae}\ee
we can Taylor expand the Lorentzian in the momentum,
\be 
p_{\rm r}\Gamma\rho\approx F_{\rm e}(t) +\gamma_e(t) p,\label{Eq:LorentzianLin3}\ee
with 
\be F_{\rm e}(t)=p_{\rm r}\Gamma \rho\left[\Delta_{\rm e}(t)\right],\quad \gamma_{\rm e}(t) = - p_{\rm r}\Gamma k \rho'\left[\Delta_{\rm e}(t)\right].\label{Eq:Fegammae0}\ee

Let us assume that $F_{\rm e}(t)$ and  $\gamma_{\rm e}(t)$ can be averaged in time independently of the torus averaging and of the spontaneous emission waiting time. This can be justified in the small amplitude limit as in \app{Sec:Final}. We define
\be \bar{F}_{\rm e} = \frac{1}{\pi}\int_0^\pi{F}_{\rm e}(t)dt, \qquad \bar{\gamma}_{\rm e} = \frac{1}{\pi}\int_0^\pi {\gamma}_{\rm e}(t)dt\label{Eq:barFe}.\ee
Then the resulting distribution will take again an exponential form
\be P(I)=\lambda_{\rm e} e^{-\lambda_{\rm e} I},\qquad \lambda_{\rm e}=-2\bar{\gamma}_{\rm e}/h_{\rm e}\ee
where 
\be \qquad h_{\rm e} = p_{\rm r} \bar{F}_{\rm e} (1+\mu)c_z.\ee
The mean final action is $\langle I\rangle=\lambda_{\rm e}^{-1}$. We now show that it can be deduced by simply measuring the photon scattering rate. Using the expansion of \eq{Eq:LorentzianLin3} we can write for the torus average defined in \eq{Eq:TorusAverageDef1D} 
\be \overline{\Gamma\rho}\approx \frac{1}{p_{\rm r}}\overline{ \left(F_{\rm e}(t) +\gamma_e(t) p\right)}=\bar{F}_e / p_{\rm r},\ee
since \eq{Eq:transf} implies that $\overline p=0$. We therefore have using \eq{Eq:meanR},
\be \langle R\rangle\approx\int dIP(I) \bar{F}_e / p_{\rm r}= \bar{F}_e / p_{\rm r},\ee
and also
\be \partial_\Delta \langle R\rangle\approx\partial_\Delta \bar{F}_e / p_{\rm r}= \overline{\Gamma\rho'[\Delta_e(t)]}= -\bar{\gamma}_e / (p_{\rm r}k).\ee
Hence we can derive the very useful relation
\be \langle I\rangle=\lambda_e^{-1}\approx\frac{1}{2}\hbar (1+\mu)\nu_z^{-1} \left[\partial_\Delta \ln \langle R\rangle\right]^{-1}.\label{Eq:Ioptimalgrad}\ee
Using \eq{Eq:Ioptimalgrad} we can deduce that the optimal detuning curve $\Delta_{\rm optimal}(A_z)$ can be found simply by looking for a maximum of the experimental observable $\partial_\Delta \ln \langle R\rangle$, and the mean action or phonon number of the resulting thermal distribution can be deduced, and in particular we obtain immediately \eq{Eq:noptimalgrad} with the proportionality constant
\be \tilde{n}= \frac{1}{2} (1+\mu)\nu_z^{-1} .\label{Eq:ntilde}\ee

\section{Thermal and nonthermal distributions in the final stage}\label{Sec:Distributions}

For the numerical calculations presented in this section we fix the Mathieu parameters for the $z$ motion,
\be a_z=-0.0002,\qquad  q_z\approx 0.16,\label{Eq:Mathieuparams}\ee
and the Floquet characteristic exponent is
\be \nu_z\approx 0.1126.\ee 
In the following subsections we consider the cases of the micromotion frequency being either low or high with respect to (half) the excited level linewidth, whereas the intermediate regime has been laid down in \seq{Sec:Intro}.

\subsection{The cooling limit with low-frequency excess micromotion}\label{Sec:Low}

In this subsection we take a $^{24}{\rm Mg}^+$ ion.
The dimensional laser parameters that we consider are 
 \be \tilde{k}\approx 2\pi / 280\,{\rm nm}^{-1},\qquad \tilde{\Gamma}\approx 263\times 10^6\,{\rm s^{-1}}\label{Eq:LaserParams1},\ee
and the spontaneous emission coefficient is
\be \mu=2/5.\label{Eq:LaserParams2}\ee
The micromotion frequency taken here is
\be 
 \Omega=2\pi\times 20\,{\rm MHz},\label{Eq:scaleparams}\ee
and the resulting secular frequency [\eq{Eq:omegaz}] is
\be \omega_z\approx 2\pi\times 1.1\,{\rm MHz}.\ee 
The condition of \eq{Eq:LowOmega} is (approximately) obeyed, and hence we are in this subsection approaching the limit of a micromotion 
 frequency much smaller than (half) the excited level decay rate.

 \begin{figure}
\includegraphics[width=3.3in]{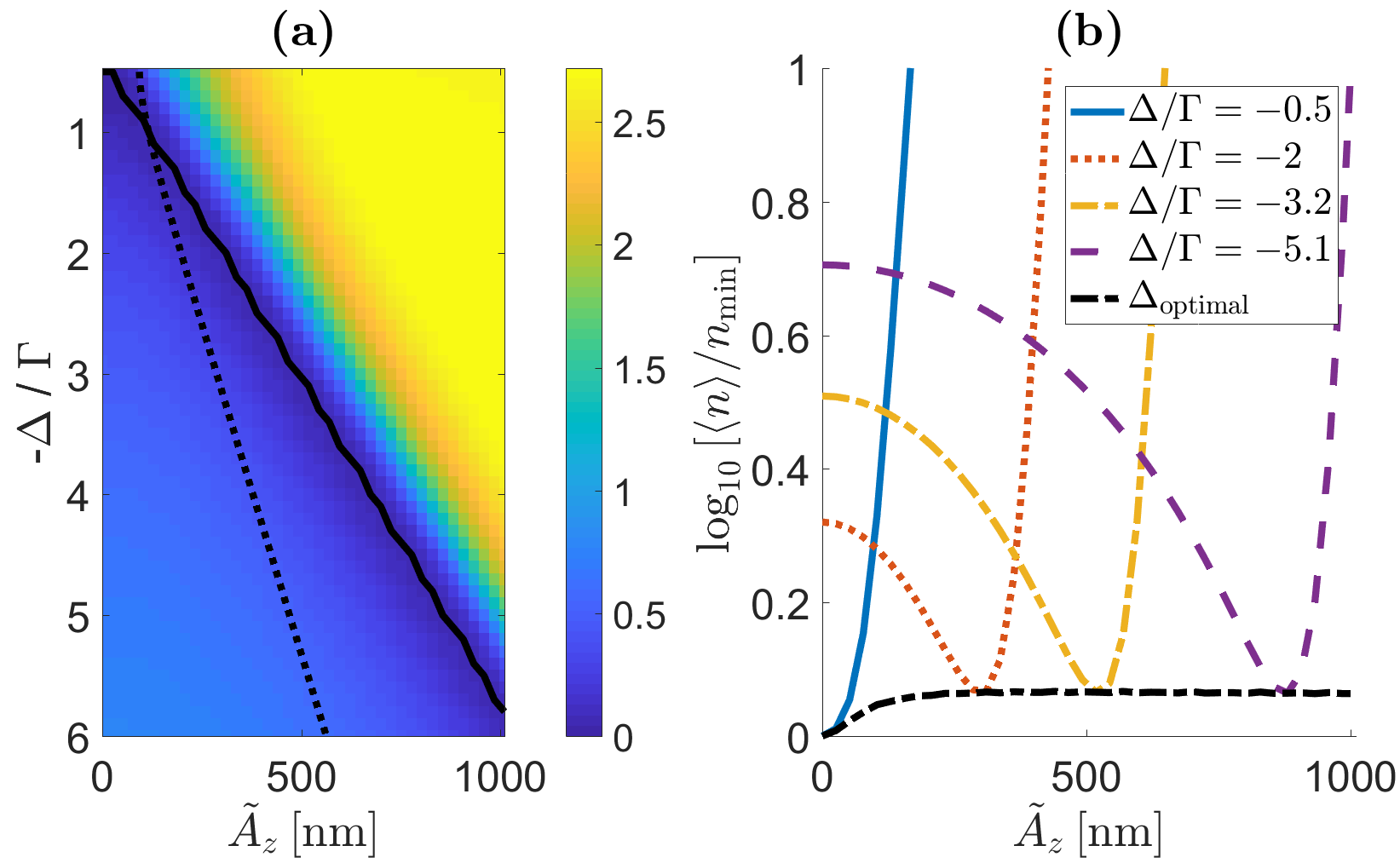}
\caption{(a) The mean final phonon number $\langle n \rangle$ [\eq{Eq:meann}] as a function of the detuning and the excess micromotion amplitude as in \fig{fig:Imean2} with the parameters of \eqss{Eq:Mathieuparams}{Eq:scaleparams}. This figure presents a calculation in the limit of low micromotion frequency, for which the photon absorption probability is taken to be the Lorentzian $\rho={\sigma}_{ee}^0$ of \eq{Eq:rho_p_z}, assuming that the micromotion dynamics are frozen on the scale of the excited level dynamics. For $ A_z \ll A_{\rm linear}(\Delta)$ [shown as the dotted curve and defined in \eq{Eq:Alinear}], the effect of the excess micromotion can be neglected. The solid black curve gives a simplified approximation [\seq{Sec:Low}] for the optimal detuning curve $\Delta_{\rm optimal}(A_z)$ for which $\langle n \rangle$ can be reduced to a minimum. (b) Curves obtained from the map of panel (a), showing the attainable minima of the action as a function of the excess micromotion amplitude. Here the vertical axis is truncated at 1, corresponding to $\langle n \rangle=10\times n_{\min}$. See the text for a discussion of the nonmonotonic features of these curves, and compare with \fig{fig:Imean31}. }
\label{fig:Imean30}
\end{figure}

Figure \ref{fig:Imean30} shows the mean phonon number $\langle n \rangle(A_z,\Delta)$ in the steady state of cooling as a function of the excess micromotion and the detuning, obtained by calculating the action distribution and using the approximate semiclassical relation \eq{Eq:napprox}, as in \eq{Eq:meann}. We here employ for the photon absorption probability the Lorentzian $\rho={\sigma}_{ee}^0$ of \eq{Eq:rho_p_z} and substitute the velocity $v_z$ of \eq{Eq:vzexpand}, which includes the excess micromotion, approximated as being frozen during the timescale of the internal level dynamics. In the figure we use the dimensional variable
\be \tilde{A}_z=wA_z,\label{Eq:tildeAz}\ee given in nanometers.
With $w=100\mu$m, $A_z$ extends to $ 10^{-2}$ in this figure, which corresponds to $\tilde{A}_z= 1000\,$nm, obtained for $E_z= 80\,$V/m.

We define the minimal mean action that is obtained in the upper left corner of the figure,
\be I_{\min}\equiv \langle I\rangle(A_z=0,\Delta=-\Gamma/2),\ee 
and using \eq{Eq:Ilimit}, we have to an accuracy of order $q_z^2/4$, $I_{\min}=I_{\rm limit}(\Delta=\Gamma/2)$.
 For increased values of $A_z$ (with $\Delta=-\Gamma/2$ fixed), the value of $\langle I \rangle$ grows slowly at first and then rises sharply, and the distribution becomes nonthermal (as will be discussed in the following). 
We show [\app{Sec:Excess}] that indeed the linear term in the expansion of $\langle I \rangle$ as a function of $A_z$ (at any fixed $\Delta$) vanishes [for the model of \eq{Eq:Mo}, with a static stray electric field]. In the region $ A_z \ll A_{\rm linear}(\Delta)$ defined by 
\be A_{\rm linear}(\Delta)= \left[\Gamma^2 +4\Delta^2\right]/(8k|\Delta|),\label{Eq:Alinear}\ee
 the effect of the micromotion is small, and $\langle I \rangle$ grows approximately inversely as a function of $\Delta$. This curve, $A_{\rm linear}(\Delta)$, is indicated by a dotted line in \fig{fig:Imean30}(a). 

 \begin{figure}
\includegraphics[width=3.3in]{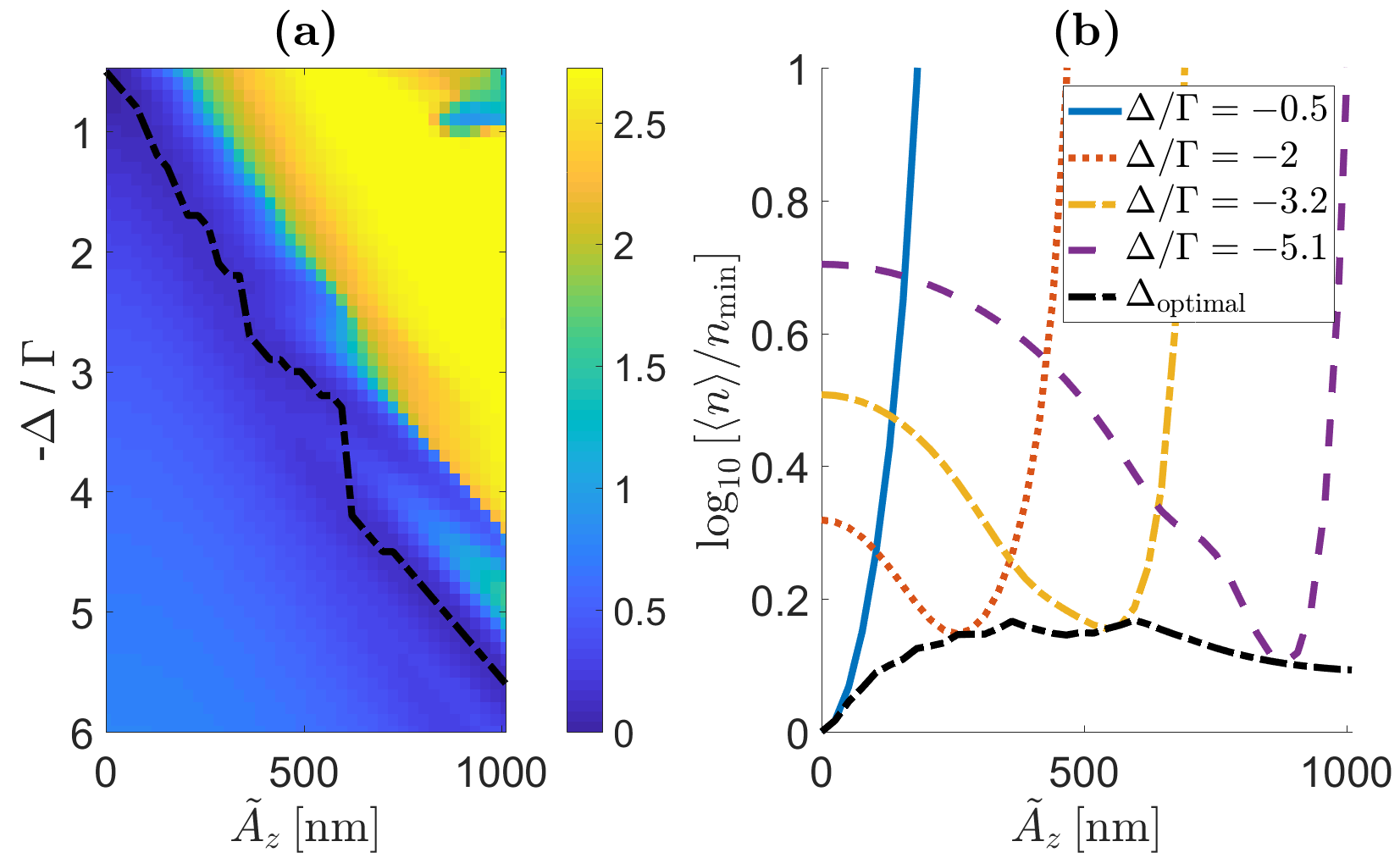}
\caption{As \fig{fig:Imean30}, but rather calculated using  the Floquet-approach photon absorption probability $\rho=\bar{\sigma}_{ee}$ of \eq{Eq:sigma0bar}, which accounts more accurately for the periodically-driven excess micromotion oscillation. The dashed-dotted black curve denotes the optimal detuning curve $\Delta_{\rm optimal}(A_z)$. In comparison with \fig{fig:Imean30}, some deviations are seen to result from narrow Floquet tongues, which are similar (but thinner than) those apparent in \fig{fig:Imean2} where the micromotion frequency is higher and comparable to the internal decay rate.}
\label{fig:Imean31}
\end{figure}

For detunings $\Delta\lesssim -0.8\Gamma$, we find that $\langle I \rangle$ first decreases with $A_z$, reaching a minimal value before growing sharply. This is a result of the coefficient at order $(A_z)^2$ in the expansion of $\langle I \rangle$ being negative in that region, as shown in \app{Sec:Excess}. 
Using the analytic form of $\rho$ we can obtain the function $\Delta_{\rm optimal}(A_z)$, by expanding the Loretzian in $p$ [see \app{Sec:Optimal}] and maintaining the periodic excess micromotion part,
\be 
p_{\rm r}\Gamma\rho\approx F_{\rm e}(t) +\gamma_e(t) p,\ee
with $F_{\rm e}(t)$ and $\gamma_e(t)$ defined in \eq{Eq:Fegammae0} taking the explicit form 
\be F_{\rm e}(t)=\frac{p_{\rm r}\Gamma s/2}{1 +\left(2\Delta_{\rm e}(t)/\Gamma \right)^2 },\quad \gamma_{\rm e}(t) = \frac{4k p_{\rm r}s \Delta_{\rm e}(t)/\Gamma}{\left[1 +(2\Delta_{\rm e}(t)/\Gamma)^2\right]^2},\label{Eq:Fegammae2}\ee
and $\Delta_{\rm e}(t)$ is defined in \eq{Eq:Deltae}.
In the absence of micromotion, the above expansion reduces to the coefficients in \eq{Eq:F0gamma1D}. The coefficients in \eq{Eq:Fegammae2} can be explicitly integrating using $\dot{\bar{z}}$ of \eq{eq:zt} approximated as in \eq{Eq:dotbarzapprox} to obtain $\bar{F}_{\rm e}$ and $\bar{\gamma}_e(t)$ of \eq{Eq:barFe} and the explicit form of the action distribution along $\Delta_{\rm optimal}(A_z)$ found from minimizing the resulting mean action. $\Delta_{\rm optimal}(A_z)$ calculated in that way can be seen to be a nearly straight line (curving only close to $\Delta=-\Gamma/2$) depicted by the solid black curve in \fig{fig:Imean30}(a), and the final action depends only weakly on $A_z$. The more accurate calculation based on the Floquet-approach photon absorption probability $\rho=\bar{\sigma}_{ee}$ of \eq{Eq:sigma0bar}, which accounts explicitly for the periodically-driven excess micromotion, is presented in \fig{fig:Imean31}, with some small but noticeable deviations (more so in the large detuning and $\tilde{A}_z$ regime). The Floquet corrections decrease in the limit of $\Omega/2\pi\ll\Gamma/2$.

 \begin{figure}
\includegraphics[width=3.3in]{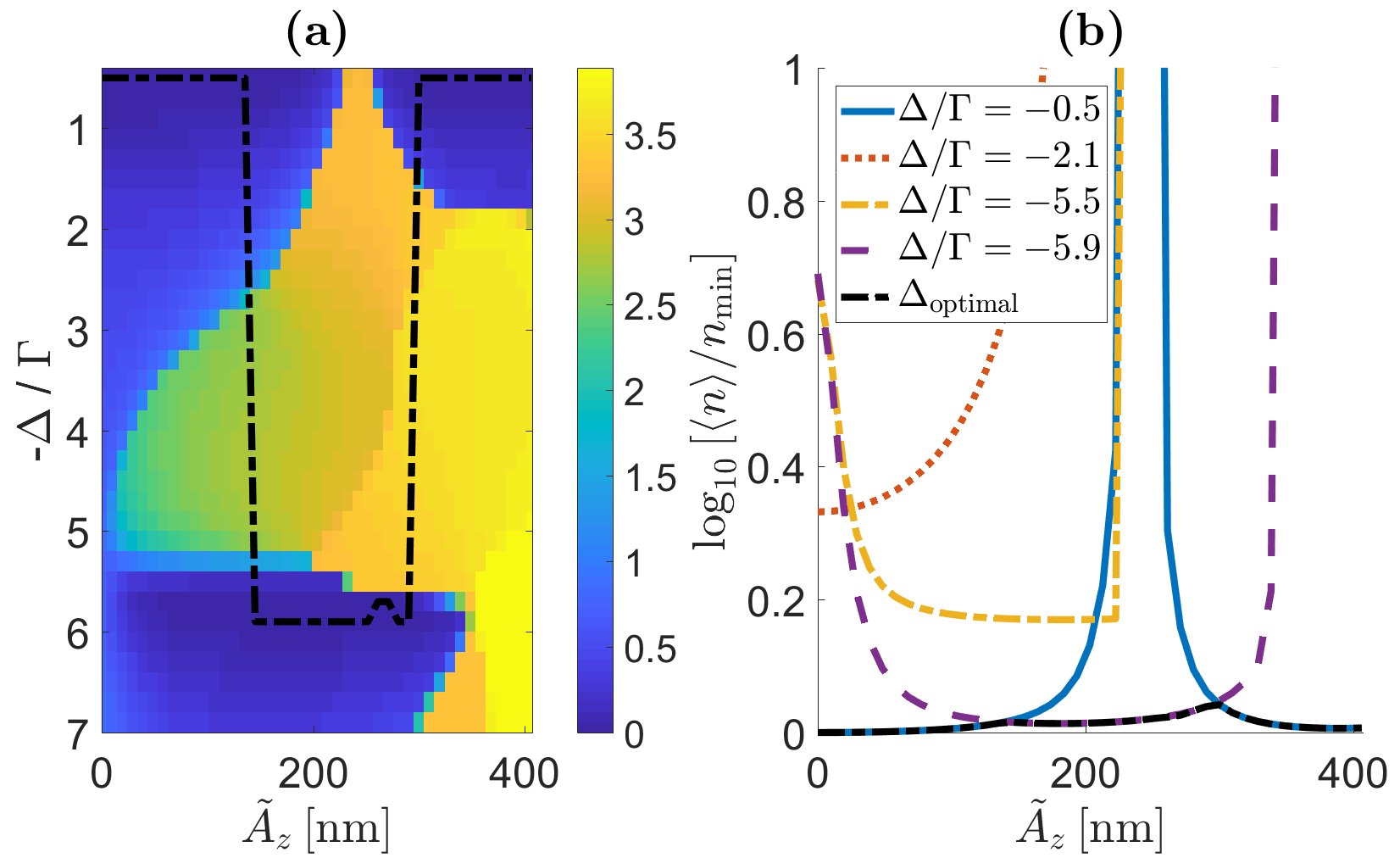}
\caption{As in \fig{fig:Imean31}, however for a $^{9}$Be$^+$ ion, with trap and micromotion parameters given in \eqss{Eq:LaserParams3}{Eq:omegaz3}, approaching the limit of a high micromotion frequency. The effect of a single Floquet resonance is clearly seen, emanating from the parametric Floquet resonance at $\Delta = -\Omega$ (the micromotion sideband), i.e.~$-\Delta/\Gamma\approx 5.3$, with a steeper dependence of the phonon distribution mean on the parameters [see panel (b)]. }
\label{fig:Imean1}
\end{figure}

\subsection{The cooling limit with high-frequency excess micromotion}\label{Sec:High}

 \begin{figure}
\includegraphics[width=3.3in]{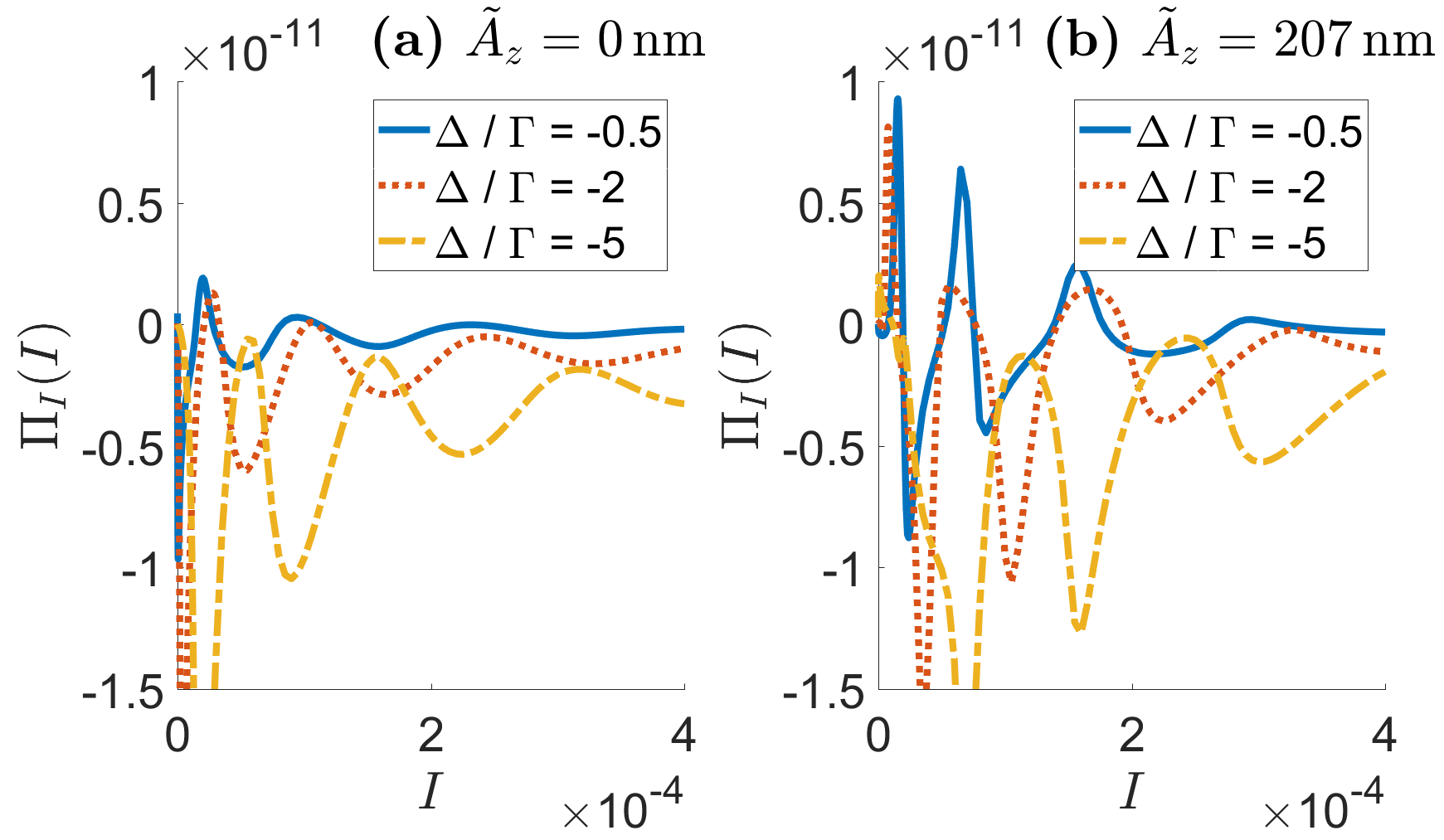}
\caption{The action drift coefficient $\Pi_I(I)$ of \eq{Eq:PiIc1D} as a function of the nondimensional action $I$, for parameters as in \fig{fig:Imean1} (with a high micromotion frequency) and using the length scale of \eq{Eq:w1}. (a) For $\Delta=-\Gamma/2$ and in the absence of excess micromotion, $\Pi_I(I)$ crosses 0 with a negative slope at two high action values, which become possible metastable points of the cooling where the ion may get captured above the steady state at the Doppler cooling limit. (b) For a large value of excess micromotion amplitude the number of such metastable zero crossings may increase, together with the width of the region where $\Pi_I(I)>0$ (which forms an effective barrier for the diffusion of the ion to the low action states). The $I$ axis extends here to motion (expanded about the excess micromotion) with an amplitude $\tilde{u}\sim 4\mu$m, and beyond the presented range the oscillations in $\Pi_I(I)$ decay and we find the asymptotic scaling $\Pi_I(I)\propto -1/\sqrt{I}$ \cite{rfcooling}. }
\label{fig:PiI4}
\end{figure}

Figures \ref{fig:Imean2}-\ref{fig:PI2} of \seq{Sec:Intro} present the analysis of the cooling with a $^{24}$Mg$^+$ ion but with a micromotion frequency which is 2.5 times larger than that taken in \fig{fig:Imean31}. As discussed briefly in \seq{Sec:Intro}, the noticeable Floquet ``tongues'' seen \fig{fig:Imean2} can be attributed to Floquet (parametric) resonances (see also \app{Sec:Bessel})). A more extensive treatment of these Floquet resonances is  beyond the scope of the current work. However,
in order to approach the limit of a high micromotion frequency with respect to the linewidth, we consider here also the example of a $^{9}$Be$^+$ ion, with
 \be \tilde{k}\approx 2\pi / 313\,{\rm nm}^{-1},\qquad \tilde{\Gamma}\approx 120\times 10^6\,{\rm s^{-1}}\label{Eq:LaserParams3},\ee
$\mu=2/5$ as in \eq{Eq:LaserParams2} and the micromotion frequency 
\be 
 \Omega=2\pi\times 100\,{\rm MHz},\label{Eq:scaleparams3}\ee
which gives using the same Mathieu parameters as in above, the secular frequency 
\be \omega_z\approx 2\pi\times 5.6\,{\rm MHz}.\label{Eq:omegaz3}\ee 
  Figure \ref{fig:Imean1} shows the structure induced by a single dominant Floquet tongue that emerges from $\Delta = -\Omega$, i.e.~$-\Delta/\Gamma\approx 5.3$. Passing this resonance corresponds to cooling on a micromotion sideband, which, as seen, can be advantageous in a range of $\tilde{A}_z$ values. 
  
  Finally, the mean steady state phonon number (or action) is not the only interesting quantity when studying laser cooling. For example, the time required to cool down from a high amplitude motion, and the probability for the ionto  be lost from the trap in the process due to diffusion, are just two examples of dynamics of the cooling process that are very interesting and can be calculated. The latter question has been treated in some detail in \cite{rfcooling}, and the former requires a detailed study of the FP dynamics. One important feature of the cooling dynamics in that case is presented in \fig{fig:PiI4}.
For comparison with the calculations presented in \cite{rfcooling} we take the distance scaling unit \be w=50\,\mu{\rm m},\label{Eq:w1}\ee
and present the action drift coefficient $\Pi_I(I)$ of \eq{Eq:PiIc1D} in terms of the nondimensional action $I$. It can be seen that for this high micromotion frequency, even without excess micromotion, there are for $\Delta=-\Gamma/2$ two high action values where $\Pi_I(I)$ crosses 0 with a negative slope, which is a possible metastable point of the cooling where the ion may get captured en route to being cooled to steady state at the Doppler cooling limit. The asymptotics of the drift coefficient beyond the presented range converges to $\Pi_I(I)\propto -1/\sqrt{I}$ as discussed in \cite{rfcooling}. For large excess micromotion the number of such metastable zero crossings may increase, together with the width of the region where $\Pi_I(I)>0$ (which forms an effective barrier for the diffusion of the ion to the low action states). For larger detunings the opposite trend is seen, with the peaks wherein $\Pi_I(I)>0$ decreasing, leading to an expected more efficient and faster cooling of the ion, strengthening the conclusion of \cite{rfcooling}; superimposing a laser with a larger detuning in traps operated at high micromotion frequency is advantageous for keeping the ion trapped and preventing it from being stuck in intermediate metastable states of motion.

\section{Summary and Outlook}\label{Sec:Outlook}

The main result of the current work has two aspects. On the practical side, our main result is that excess micromotion of a large amplitude does not prevent reaching a low, thermal-like distribution for the fluctuations expanded about the coherent driven motion, achievable by simply tuning the cooling laser frequency. This is useful in particular when considered in the context of ion crystals discussed below. This concrete result also demonstrates a broad conclusion that can be drawn, which is that even though the ion is strongly driven by the trap potential in combination with the stochastic photon scattering processes, by choosing the right frame, the driven and the stochastic dynamics can be separated, with each becoming considerably simpler. The action-angle coordinates form the right coordinates because the fluctuations in action represent the relevant stochastic, noncoherent part of the motion, and moreover the angles can be safely coarse grained, leaving a clear picture of the underlying complex dynamics in terms of the actions.

An important foreseeable application of the presented results is to a crystal of ions. In a crystal configuration for which some ions are not positioned at a point where the rf potential vanishes (e.g.~in a planar configuration in a linear Paul trap), those ions will perform driven periodic motion akin to excess micromotion, determined by the interplay of the Coulomb interaction and the periodic drive \cite{rfions}. It is the dynamic equivalent of an ion's equilibrium position in a static crystal, and cannot be removed.
By tailoring the laser detuning, the final action distribution of these ions can be significantly reduced.
 A setup consisting of a few lasers can be treated by adding the drift and diffusion coefficients calculated separately for each laser. Spatially inhomogeneous laser profiles and laser parameters which are modulated in time can be transparently treated. In combination with noise heating \cite{rfcooling}, a detailed characterization of the final distribution of laser-cooled ions and the dynamics leading to it can be obtained, and ideas for its manipulation can be explored \cite{chupeau2018engineered}.
 The ensuing dynamics can be studied using the analytic tools of \app{Sec:Mathieu}, by expansion in small linearized deviations about the driven periodic motion.
This way, cooling and heating dynamics corresponding to Gaussian white noise, and the stationary distribution of chains of ions in 1D and crystals in 2D and 3D configurations, can be studied \cite{PhysRevA.64.063407, morigi2003ion,morigi2001twospecies, fogarty2016optomechanical,laupretre2018controlling, kamsap2017experimental, PhysRevLett.119.043001,1367-2630-13-4-043019, PhysRevX.8.021028,Mitchell13111998, Drewsen_Long_Range_Order,
SchilerProteinsPRL,szymanski2012large,tabor2012suitability, mavadia2013control,bohnet2016quantum}. The theory of  Ornstein--Uhlenbeck processes can be applied for a detailed analysis and at the same time be tested experimentally in a controlled way \cite{godreche2018characterising}.

Interesting extensions of the theory could include more general electronic level structures \cite{PhysRevA.96.012519,janacek2018effect}, and applying the action-angle framework to setups where power-law distributions in energy (in an averaged sense) were predicted for collisions of ions with neutral atoms \cite{devoe2009power,rouse2017superstatistical, rouse2018energy,meir2018direct}.
The interplay of micromotion, noise and laser cooling is of significant importance for applications in quantum information processing and the operation of quantum gates and entanglement operations with trapped ions  \cite{PhysRevLett.81.3631,berkeland1998minimization,PhysRevLett.101.260504, rfions,zigzagexperiment, Landa2014,PhysRevA.90.022332,wang2015quantum, arnold2015prospects, keller2015precise,yan2016exploring, mielenz2016arrays, bruzewicz2016scalable, keller2017optical,keller2018controlling, welzel2018spin,delehaye2018single,PhysRevA.97.062325}. In particular, as discussed above,
the actions standing at the heart of the current work correspond exactly to the quantum mechanical phonons with a periodically-driven harmonic potential \cite{Glauber1992}, in terms of Floquet-Lyapunov modes \cite{rfions,rfmodes}.
Using the exact time-dependent wavefunctions in quadrupole traps \cite{rfmodes,mihalcea2017study}, would allow to extend the theory to the quantum limit.


\begin{acknowledgments}
H.L. thanks Ananyo Maitra, Dietrich Leibfried, Denis Ullmo, and Roni Geffen for fruitful discussions, and acknowledges support by IRS-IQUPS of Universit\'{e} Paris-Saclay, and by LabEx PALM under grant number ANR-10-LABX-0039-PALM.

\end{acknowledgments}

\appendix

\section{A coupled system of Mathieu oscillators}\label{Sec:Mathieu}

We consider an  ion that has been cooled to the center of a Paul trap of a general type. We assume that the potential can be approximated as a quadrupole (this is however not necessarily the case in multipole traps \cite{MartinaRing}). 
With $\vec{r}$ and $\vec{v}\equiv \dot{\vec{r}}$ being the vector coordinate and velocity of an ion in $D=3$ dimensions, we first rescale the time $t$ by half the micromotion frequency, and the coordinates by a natural unit of length, $w$, relevant for the trap at hand,
\begin{equation}
 \vec{r} \to \vec{r}/w, \qquad t \to \Omega t /2,\qquad  \vec{v} \to \vec{v}/(w\Omega/2).
\label{Eq:rescaling}
\end{equation}
Using the rescaling in \eq{Eq:rescaling} allows us also to define a nondimensional momentum using the ion mass $m$ and by absorbing it and the ion charge $e$ into the parameters, we define a nondimensional potential energy $V$ resulting from an electrostatic voltage $U$,
\be V \to V / [ mw^2\Omega^2/(4e)],\ee
a nondimensional electric field,
\be E \to E / [mw\Omega^2/(4e)],\ee
and a nondimensional Planck constant scaled according to
\be \hbar\to {\hbar} /(m w^2\Omega/2).\label{Eq:hbar}\ee

In general, the quadrupole potential may be composed of a sum of a few quadrupole potential terms whose origin does not coincide, or there may be ``stray'' electric fields that push the ion from the origin of the quadrupole potential. The resulting motion contains a component known as ``excess micromotion'', since it can (typically) be minimized by using additional DC electric fields. 
In this case it is useful to describe the motion by using generalized coordinates $\vec{u}$ with conjugate momenta $\vec{p}$, differing from the real space position $\vec{r}$ and velocity $\vec{v}$ (with $m=1$) by a time-dependent displacement \cite{rfions}, that eliminates the terms linear in the coordinates from the potential energy,
 \be \vec{r}(t)=\bar{r}^0(t)+ \vec{u}(t), \qquad \vec{v}(t)=\dot{\bar{r}}^0(t)+ \vec{p}(t).\label{Eq:r0}\ee
We assume that the nondimensional potential $V(\vec{u},t)=V(\vec{u},t+\pi)$ is time-reversal invariant and $\pi$-periodic (which can include the particular case where it is time-independent), and can be expanded into a system of $D$ coupled Mathieu oscillators \cite{rfions}, obtaining the nondimensional Hamiltonian 
\be
H_0 (\vec{u},\vec{p},t)=\frac{1}{2}(\vec{p})^2+V_{\rm M.o.}(\vec{u},t),
\ee
with
\be V_{\rm M.o.}(\vec{u},t)=\frac{1}{2}\vec{u}^t\left(A-2Q\cos 2t\right)\vec{u}\label{Eq:VMo}
\ee
where $\vec{u}^t$ denotes the transpose, and $A$, $Q$ are matrices that describe the linearized DC and rf voltages tensors. By the Laplace equation, ${\rm tr} A={\rm tr} Q=0$. The equations of motion derived from the linearized potential form a coupled system of parametric oscillators \cite{rfmodes,shaikh2012stability}.
If $A$ and $Q$ commute, then they can be diagonalized to give a system of decoupled Mathieu equations. In the opposite case, no such simple transformation exists, and the three spatial directions will be mixed by the micromotion.

The most general solution of this motion is a sum over $D$ decoupled linear oscillators
\begin{align}
&\vec{u}= \sum_j \sqrt{I_j}\left(2{\rm Re}\sum_n \vec{C}_{2n}^j e^{i2nt}e^{i\theta_j}\right),\label{Eq:uMathieu}\\
& \vec{p}= -\sum_j \sqrt{I_j}\left(2{\rm Im}\sum_n \vec{C}_{2n}^j (2n+\nu_j)e^{i2nt}e^{i\theta_j}\right),\label{Eq:pMathieu}
\end{align}
where the $n$ summation extends over $\mathbb{Z}$ and the coefficients $\vec{C}_{2n}^j $ for $n\neq 0$ give the micromotion modulation. In \eqs{Eq:uMathieu}-\eqref{Eq:pMathieu} we have defined
\be \theta_j=\nu_j t+\phi_j,\label{Eq:thetadef}\ee
 with $\nu_j$ the characteristic exponents of the Mathieu system, that are related to the secular frequencies of motion in the trap ($\omega_j$, in physical units), by
\be \omega_j =  \nu_j \Omega /2,\label{Eq:omegaj}\ee
and ${I_j}$ and $\phi_j$ are (for now) arbitrary constants related to the initial conditions. 
We assume that the motion is stable, i.e.~that $\nu_j$ are all real, and are between 0 and 1.

As detailed in \cite{rfmodes}, a time-dependent (Floquet-Lyapunov) linear transformation can be used to transform the real space coordinates and momenta, to new (complexified) coordinates $\vec{\xi}$ and the canonically conjugate momenta $-i\vec{\chi}$,
\be \left(\begin{array}{c} {\vec{u}} \\ {{\vec{p}}} \end{array}\right)=\Gamma(t) \left(\begin{array}{c} {\vec{\xi}} \\ {{\vec{\chi}}} \end{array}\right),\qquad \Gamma \left(t\right)=\left(\begin{array}{cc} {U} & { U^*} \\ {V } & {V^* }\end{array}\right)\label{Eq:FL}\ee
where and $U(t)$, $V(t)$ are $D\times D$ complex matrices constructed from the $\pi$-periodic part of the $D$ column vector solutions given in \eqs{Eq:uMathieu}-\eqref{Eq:pMathieu}, i.e.
\be U = \left(\sum \vec{C}_{2n}^j e^{i2nt}\, ...\right), V = \left(i\sum (2n+\nu_j)\vec{C}_{2n}^j e^{i2nt}\, ...\right),\ee
and $U^*$ is the complex conjugate matrix, and we will use $U^\dag$ for the hermitian conjugate matrix, and also $U^t$ that is the transposed matrix.
These matrices are to be rescaled by multiplication by a constant diagonal matrix,
\be U\to U(-2i V^t(0)U(0))^{-1/2},\label{Eq:Urescale}\ee
and $V$ similarly, 
guaranteeing the normalization
\be V^t(0)U(0)=\frac{1}{2}i,\label{Eq:VtUNormalization}\ee
which also implies the identity (that will be used later),
\be U^t(t)V^*(t)-V^t(t)U^*(t)=-i.\label{Eq:UtVsIdentity}\ee

Then the transformation in \eq{Eq:FL} is canonical and its inverse is
\be \left(\begin{array}{c} {\vec{\xi}} \\ {{\vec{\chi}}} \end{array}\right)=\Gamma^{-1}(t) \left(\begin{array}{c} {\vec{u}} \\ {{\vec{p}}} \end{array}\right),\quad \Gamma^{-1} \left(t\right)=\left(\begin{array}{cc} {iV^\dag} & { -iU^\dag} \\ {-iV^t } & {iU^t }\end{array}\right).\label{Eq:FLinv}\ee
The Hamiltonian then transforms according to 
\be H_0(\vec{\xi},\vec{\chi}) = \frac{1}{2}\sum_j\nu_j\left(\xi_j \chi_j+\chi_j \xi_j\right).\ee
The time evolution of the new canonical coordinates is given by (noting that $\xi_j=\chi_j^*$)
\be {\xi}_j\left(t\right)= \sqrt{I_j} e^{i(\nu_jt+\phi_j)}, \qquad
{\chi}_j\left(t\right)= \sqrt{I_j}e^{-i(\nu_jt +\phi_j)}. \label{Eq:OperatorsSolution}\ee 

We can now introduce naturally a second canonical transformation to the action-angle coordinates $(\vec{I},\vec{\theta})$,
\be I_j=\xi_j \chi_j,\qquad \theta_j=i\ln\chi_j,\label{Eq:Ixichi}\ee 
which justifies the notation in \eq{Eq:thetadef}, since
\be {\xi}_j= \sqrt{I_j} e^{i\theta_j}, \qquad
{\chi}_j= \sqrt{I_j}e^{-i\theta_j}, \label{Eq:xizetaItheta}\ee 
whence the Hamiltonian obtains the simple form
\be H_0(\vec{I})=\sum_j \nu_jI_j.\ee
Since this is the canonical Hamiltonian of simple decoupled harmonic oscillators, the quantum levels can be obtained by using the well-known relation semiclassical quantization relation \cite{bohigas1993}, \be n_j+1/2=I_j/\hbar,\label{Eq:quantization}\ee
corresponding exactly to the precise wavefucntions \cite{rfmodes}.
The relations in \eq{Eq:Ixichi} define implicitly the action-angle transformation from the real-space coordinates,
\begin{align}
&I_j=\Lambda_j(\vec{u},\vec{p},t), && \theta_j=\Theta_j (\vec{u},\vec{p},t),\label{Eq:AAtransfo}\end{align}
with the transformation functions $\Lambda_j$ and $\Theta_j$  depending explicitly on time (and being $\pi$-periodic), and 
can be constructed explicitly using \eq{Eq:Ixichi} and \eq{Eq:FLinv}. In the following we will not need the functions $\Lambda_j$ and $\Theta_j$ explicitly, however we will use the partial derivatives
\be \frac{\partial \Lambda_j}{\partial p_\alpha}=iU_{\alpha j}\xi_j-iU^*_{\alpha j}\chi_j,\quad \frac{\partial^2 \Lambda_j}{\partial p_{\alpha}\partial p_{\beta}}= U_{\alpha j} U^*_{\beta j} + U_{\alpha j}^* U_{\beta j}.\label{Eq:LambdaDerivatives}\ee

We note here that the entire derivation above would remain completely valid if \eq{Eq:VMo} is to  be replaced by a more general time-reversal invariant and $\pi$-periodic linear system, i.e.~a system of coupled Hill equations \cite{McLachlan,Yakubovich,rfions,rfmodes}, with higher harmonics of the fundamental frequency $2$. In addition, we note that setting $\vec{C}_{2n}^j=\delta_{n,0}\vec{C}_{0}^j /(2\nu_j)^{1/2}$, the solutions for $\vec{u}$, $\vec{p}$ reduce to coupled  harmonic oscillators with nondimensional frequencies $\nu_j$. 
This is the pseudopotential approximation (within the harmonic approximation), with the motion described by the potential \be V_{\rm h.o.}(\vec{u})=\frac{1}{2}\vec{u}^t N\vec{u}, \label{Eq:Vho}\ee
with $N$ a time-independent coupling matrix.

\section{The linear limit of cooling for a coupled Mathieu system}\label{Sec:Final}

As derived in \cite{rfcooling}, a Fokker-Planck equation in $D$ actions (equal to the space dimension)  can be written using the probability flux vector $\vec{S}$ whose components are given by
\begin{align}
{S}_j(\vec{I},t)  \equiv  \Pi_j P - \frac{1}{2}\sum_k \frac{\partial }{\partial I_k} \cro{\Pi_{jk} P},
\label{Eq:FPS}
\end{align}
with the FP equation taking the form
\begin{align}
&\frac{\partial {P}(\vec{I},t) }{\partial t} = -\sum_j\frac{\partial {S_j}(\vec{I},t) }{\partial I_j} = \nonumber \\& - \sum_j\frac{\partial }{\partial I_j} \cro{\Pi_j{P}} + \frac{1}{2}\sum_{j,k}\frac{\partial^2 }{{\partial I_j}\partial I_k} \cro{\Pi_{jk}{P}}.
\label{Eq:FPIt}
\end{align}
For a multidimensional torus, generalizing  \eq{Eq:Za}, a phase-space point is defined by
\be Z_a\equiv \{\vec{u},\vec{p},t\},\ee
and 
defining the torus average over any function $\Xi(\vec{I},\vec{\theta},t)$ of the phase space, where $\Xi$ is assumed to have an arbitrary period $T$,
\be \Xi\left( \vec{I},\vec{\theta},t+T\right)=\Xi\left( \vec{I},\vec{\theta},t\right),\ee 
we have
\be
\overline{\Xi}\left( \vec{I}\right)\equiv \frac{1}{T} \int_0^{T}dt\frac{1}{(2\pi )^{ D}}\int \Xi\left(\vec{I},\vec{\theta},t\right) d^{ D}\vec{\theta},\label{Eq:TorusAverageDef}
\ee 
with the adiabaticity conditions, 
\be \Pi_{j}(\vec{I})/ \nu_j \ll I_j, \quad \Pi_{jk}(\vec{I})/ \sqrt{\nu_j \nu_k}  \ll I_j I_k.\label{Eq:AdiabaticCondition}\ee 
The mean drift and diffusion rates are given by 
\be
\Pi_{j}(\vec{I}) =\Gamma\overline{ \rho(Z_a)\langle\delta I_j\rangle},\quad  
\Pi_{jk}(\vec{I}) =\Gamma\overline{ \rho(Z_a)\langle  \delta I_{j}\delta I_k\rangle},\label{Eq:FPPcij}
\ee
 with
 \bem \langle\delta  I_{j} \rangle  =
p_{\rm r} \sum_\alpha {\hat{k}}_\alpha\frac{\partial
  \Lambda_j(Z_a)}{\partial p_\alpha} \\+ \frac{1}{2} p_{\rm r}^2
\sum_{\alpha,\beta}\left[  {\hat{k}}_\alpha {\hat{k}}_\beta
  \frac{\partial^2 \Lambda_j(Z_a)}{\partial p_\alpha\partial
    p_\beta}+\mu_{\alpha\beta} \left\langle\frac{\partial^2
      \Lambda_j(Z_a)}{\partial p_\alpha\partial p_\beta}
  \right\rangle_\Gamma  \right],\label{Eq:deltaI1}\end{multline} 
and  
\be \langle\delta I_{j}\delta I_k\rangle = p_r^2
      \sum_{\alpha,\beta}\left[\hat{k}_\alpha\hat{k}_\beta
        \frac{\partial\Lambda_j}{ \partial p_\alpha}
        \frac{\partial\Lambda_k}{\partial p_\beta} + \mu_{\alpha\beta}
        \left\langle\frac{\partial\Lambda_j}{ \partial p_\alpha}
          \frac{\partial\Lambda_k}{\partial p_\beta}
        \right\rangle_\Gamma \right] ,\label{Eq:Sigma2}\ee 
       where all terms on the r.h.s above are
      functions of $Z_a$.
The absorption probability for $s\ll 1$ takes the form of the Lorentzian $\rho={\sigma}_{ee}^0$
of \eq{Eq:rho_p_z}, 
\be\rho(\vec{p},t) = \frac{s/2}{1+\pare{2\Delta_{\rm{eff}}/\Gamma}^2}, \qquad \Delta_{\rm eff}=\Delta-\vec{k}\cdot \vec{v}(\vec{p},t),\label{Eq:rho_p_z2}\ee
with  $\Delta_{\rm eff}$ showing that $\vec{v}(\vec{p},t)$ can be a function of the canonical momentum and time [\eq{Eq:r0}].

Following the treatment of \app{Sec:Mathieu} we consider  motion described by the coupled Mathieu system of \eq{Eq:VMo}, and in this section we assume for simplicity that excess micromotion is negligible, i.e.~we set
\be \bar{r}^0(t)=0.\ee
 Under these conditions, that the velocities are small and that the potential is quadrupole, we can derive the following expressions for the linear limit of the cooling, which turns out to be identical within both the zero lifetime limit and the finite lifetime assumptions of the derivation.

Having assumed small velocities in this limit, we  set $\vec{v}=\vec{p}$ in the rescaled units. Then, linearizing the Lorentzian in velocity we can write the coefficients in a well-known form \cite{javanainen1980,leibfried2003,rfcooling},
\be 
p_{\rm r}\Gamma\rho(\vec{p})\approx F_{\rm r} + \gamma \sum_\beta  \hat{k}_\beta p_\beta,\label{Eq:LorentzianLin}\ee
with
\be F_{\rm r}=\frac{p_{\rm r}\Gamma s/2}{1 +(2\Delta/\Gamma)^2}, \quad \gamma = \frac{4k p_{\rm r}s \Delta/\Gamma}{\left[1 +(2\Delta/\Gamma)^2\right]^2}.\label{Eq:F0gamma2}\ee
Here $F_{\rm r}$ gives a mean radiation force (for $\vec{p}=0$), and $\gamma$ is a rate of damping. The condition for the validity of this linearization is 
\be \vec{k}\cdot \vec{v} \ll \left[\Gamma^2+4\Delta^2\right]/(8|\Delta|)\label{Eq:kvLin}.\ee

Using \eqs{Eq:FL}, \eqref{Eq:Ixichi} and \eqref{Eq:LambdaDerivatives}, the tori averages in  \eq{Eq:FPPcij} would then contain bilinear terms of the oscillators $\vec{\xi}$ and $\vec{\chi}$. 
Let us assume first that the ion can be assumed stationary in space between the absorption and emission, which can be expressed by substituting 
\be \Gamma e^{-\Gamma t'}\approx \delta(t')\label{Eq:deltadecay}\ee in $\langle \cdot\rangle_\Gamma $ of \eq{Eq:GammaDecay}. We will see below that the result without this restriction is in fact identical.
Assuming that $\{\nu_j\}$ are nondegenerate \footnote{For an analysis of laser-induced correlations between two nearly-degenerate harmonic oscillator modes, see \cite{javanainen1980}.}, the only surviving combination will involve $\xi_j\chi_j$, due to \eq{Eq:xizetaItheta}. We hence find  that these terms become constants, or linear in the actions. Defining the coefficients
\begin{align}
&c_{\alpha \beta}^j=\overbar{\left( U_{\alpha j} U^*_{\beta j} + U_{\alpha j}^* U_{\beta j}\right) }, \label{Eq:FPc}\\ &d_{\alpha \beta}^j=\overbar{\left( iU_{\alpha j} V_{\beta j}^*-iU_{\alpha j}^* V_{\beta j}\right) } \label{Eq:FPd},
\end{align}
 allows us to put the FP coefficients in the form
\be \Pi_{j}^l(\vec{I}) = g_{j} I_j+h_j/2,\qquad \Pi_{jk}^l(\vec{I}) =\tilde{ h}_{j}I_j\delta_{j,k} ,\ee
with
\begin{align}
g_{j} = {\gamma }\sum_{\alpha,\beta}\hat{k}_{\alpha}\hat{k}_{\beta} d_{\alpha \beta}^j, \qquad h_j=\tilde{ h}_j= \sum_{\alpha,\beta} D_{\alpha\beta}^lc_{\alpha \beta}^j, \label{Eq:FPgh}
\end{align}
and \be
 D_{\alpha\beta}^{l} = p_{\rm r}F_{\rm r}(\hat{k}_\alpha \hat{k}_\beta+\mu_{\alpha\beta}).\ee
We note that the fact that the same coefficient $h_j=\tilde{ h}_j$ appears in both $\Pi_j^l$ and $\Pi_{jj}^l$ after the averaging (where they result from averaging different terms), is a consequence of the linearity of the oscillations, and is important in the following. In addition, for an isotropic laser, i.e.~if $\hat{k}_{\alpha}=\hat{k}_{\beta}$, then \eq{Eq:UtVsIdentity} implies that $g_j=\gamma$. 

For a zero current state, with the probability flux vector defined in \eq{Eq:FPS}, we require
\be\Pi_j^l P=\frac{1}{2}\frac{\partial}{\partial I_j}(\Pi_{jj}^l P).\ee
Substituting the exponential ansatz 
\be P=\left(\prod_j \lambda_j\right)\exp\left\{-\sum_j \lambda_j I_j\right\},\ee
the zero-current condition becomes
\be g_{j} I_j+\frac{1}{2}h_j=\frac{1}{2}\left[\tilde{h}_j- \lambda_j \tilde{h}_j I_j\right],\ee
and here we see that in order for the constant term to cancel on both sides of the equation, we must have $h_j=\tilde{h}_j$, i.e.~the same coefficient in both the drift and the diffusion. The distribution is then solved by
\be \lambda_j=-2g_j/h_j\label{Eq:Imeanlambda}.\ee
This is a thermal-like, equilibrium distribution with mean action
\be \left\langle I_j\right\rangle = 1/\lambda_j,\ee
provided that $g_j<0$. This requires $\Delta<0$ and also that none of the normal modes decouples from the laser. 

Relaxing the instantaneous decay assumption of \eq{Eq:deltadecay}, we now carry out the integration over the waiting time (exponential) distribution for decay from the excited level. The tori averages will result in terms which can be written in the form
\begin{widetext}
\be \overbar{\left\langle \frac{\partial\Lambda_j}{ \partial p_\alpha} \frac{\partial\Lambda_k}{ \partial p_\beta}\right\rangle_\Gamma}= \sum\sqrt{I_j I_k}C^{\alpha j}_{2n}C^{\beta j}_{2m}
\int_0^\pi \frac{1}{\pi} dt_a \int_0^{2\pi}\frac{1}{2\pi} d\phi_j \int_0^{2\pi} \frac{1}{2\pi} d\phi_k \int_0^{\infty} dt'\Gamma e^{-\Gamma t'} e^{i(\pm 2n\pm 2m\pm \nu_1\pm \nu_2)(t_a+t')}e^{i(\pm \phi_j\pm \phi_k)}.\label{Eq:pLG}\ee
\end{widetext}

In the above integral, the order of integration does not matter, and hence the integration over $t'$ can be performed last. In that case, the expression in \eq{Eq:pLG} reduces, by the arguments leading to \eq{Eq:FPc}, to
\be \overbar{\left\langle \frac{\partial\Lambda_j}{ \partial p_\alpha} \frac{\partial\Lambda_k}{ \partial p_\beta}\right\rangle_\Gamma}=\int_0^{\infty} dt'\Gamma e^{-\Gamma t'} c^j_{\alpha \beta}I_j  = c^j_{\alpha \beta}I_j,\ee
and similarly,
\be \overbar{\left\langle\frac{\partial^2\Lambda_j}{ \partial p_\alpha\partial p_\beta} \right\rangle_\Gamma} = c^j_{\alpha \beta}.\ee
This change of integration order does not hold ({\it a-priori}) if higher order terms in the Lorentzian expansion of \eq{Eq:LorentzianLin} have to be included, because they will appear outside of the $dt'$ integration (the term linear in the momentum integrates to 0 in any case). 

Hence the final stage of the cooling is described by an equilibrium distribution in the action coordinates, even when the ion cannot be treated as frozen between the absorption and emission, and even if it is driven by the high frequency rf trap to a large kinetic energy. Sufficient conditions for the validity of this approximation are the linearization [\eq{Eq:kvLin}] and the nondegeneracy of the modes, in addition to the conditions of the derivation of the finite lifetime particle limit, presented in \cite{rfcooling}. 

\section{Cooling dynamics in 1D}\label{Sec:Derivations}

\subsection{Hamiltonian dynamics}\label{Sec:Hamiltonian1D}

Using the general derivation of \app{Sec:Mathieu},    the  solution of the linear and homogeneous, Mathieu oscillator equation of motion derived from \eq{Eq:H0_1D} is given by
\be
u = \sqrt{I}\left(U(t) e^{i\theta }+U(t)^* e^{-i\theta }\right),\ee
and
\be {p}= \sqrt{I}\left(V(t)e^{i\theta }+V(t)^* e^{-i\theta }\right).\label{Eq:zpMathieu}
\ee
The functions $U(t)$ and $V(t)$ 
are both $\pi$-periodic functions,
\be U(t) = \sum_n {C}_{2n} e^{i2nt},\quad V(t)= i \sum_n {C}_{2n} (2n+\nu_z)e^{i2nt},\label{Eq:UV}\ee
and the normalization of the coefficients is given by 
\be V(0)U(0)=\frac{1}{2}i.\label{Eq:VtUNormalization1D}\ee
The partial derivatives of the action angle transformation of \eq{Eq:Itheta} are given by
\bea \frac{\partial \Lambda}{\partial p}=\sqrt{I}\left(iU(t)e^{i\theta}-iU(t)^* e^{-i\theta}\right),\quad \frac{\partial^2 \Lambda}{\partial^2 p}= 2|U(t)|^2.\label{Eq:LambdaMoDerivatives}\end{align}

For the particular solution of the inhomogenous equation derived from 
 $ {V}_{\rm{e}}$ of \eq{Eq:Mo},
 we can substitute $\bar{z}$ of \eq{eq:zt}, obtaining the recursion relations
 \be \left[B_{2n}(a_z-4n^2)- q_z(B_{2n-2}+B_{2n+2})\right]=E_z\delta_{n,0}.\ee
 To get a continued fraction expansion we use the assumed time-reversal invariance of the solution, $B_{2n}=B_{-2n}$, and define $c_{2n}=B_{2n+2}/B_{2n}$, getting $c_{2n-2}c_{2n}=B_{2n+2}/B_{2n-2}$, so
 \be B_0=E_z/(a_z-2q_z c_0),\ee
 \be c_{2n-2}=q_z /(a_z-4n^2- q_z c_{2n}),\quad n\ge 1.\ee
To the leading order [\eq{Eq:nuapprox}] we have for the coefficients of $\bar{z}$,
\be B_0 \approx  E_z/\nu_z, \qquad B_2\approx -B_0 q_z /4, \qquad B_4\approx 0.\ee

\subsection{Low-Frequency Excess Micromotion}\label{Sec:Excess}

In the absence of excess micromotion, the final stage of the cooling in the Mathieu oscillator potential has been studied in \cite{rfcooling} and here, above. We now generalize this treatment to include excess micromotion in the limit of \eq{Eq:LowOmega} [with a micromotion frequency smaller than the excited state linewidth], focusing on 1D motion. 

We assume that the low-frequency micromotion Lorentzian $\rho={\sigma}_{ee}^0$
of \eq{Eq:rho_p_z} can be expanded in the velocity, subject to conditions of validity that will be elucidated later. We carry the expansion to third order in $v_z$, and then we set ${v_z}=\dot{\bar{z}}+{p}$ in the rescaled units (with the mass set to 1), and assume in addition that $p$ remains small, obtaining
\be 
p_{\rm r}\Gamma\rho\approx F_{\rm r} +\gamma v_z+ \kappa v_z^2\approx F_{\rm r} +\gamma \dot{\bar{z}}+ \kappa \dot{\bar{z}}^2 +\gamma p,\label{Eq:LorentzianLin2}\ee
with $F_{\rm r}$ and $\gamma $ defined in \eq{Eq:F0gamma1D},
and the  nonlinear damping coefficient is
\be \kappa = \frac{2k^2 p_{\rm r}s (1-12 (\Delta/\Gamma)^2)/\Gamma}{\left[1 +(2\Delta/\Gamma)^2\right]^3}.\ee

Plugging \eq{Eq:LorentzianLin2} and \eq{Eq:LambdaMoDerivatives} into \eqss{Eq:PiIc1D}{Eq:PiIIc1D}, using \eq{Eq:UtVsIdentity} and performing the averaging of \eq{Eq:TorusAverageDef} by exploiting the fact that the order of the integration of the exponential decay of \eq{Eq:GammaDecay} can be interchanged with the averaging in this limit of linearization [\app{Sec:Final}], we get
\be \Pi_I = \gamma I + j_z/2,\qquad \Pi_{II}= j_z I,\ee
where the coefficient of excess micromotion is
\be j_z= p_{\rm r}(1+\mu)\left(F_{\rm r}c_z + \gamma d_z+\kappa e_z\right),\ee
with
\be c_z = \overline{2\left|U\right|^2},\label{Eq:c_z}\ee
and
\be d_z=\overline{2\dot{\bar{z}} \left|U\right|^2},\qquad e_z=\overline{2\dot{\bar{z}}^2 \left|U\right|^2}.
\ee
We note that $j_z$ generalizes $i_z$ of \eq{Eq:FP1Dh} (for which $d_z$ and $e_z$ both vanish), which  generalizes a similar coefficient for the harmonic oscillator, $h_z$ defined in \cite{rfcooling}, for which $c_z$ is replaced by $\nu_z^{-1}$.
Since $ \dot{\bar{z}} $, given in \eq{Eq:dotbarzapprox}, is odd under time-reversal , while $|U(t)|^2 = U(t)U(t)^*$ is even (as $U(t)$ is a sum of exponentials), we see that $d_z=0$, so the leading order correction to the action distribution is in second order in the excess micromotion amplitude $A_z$ defined in \eq{Eq:Az}. We can also see from the expression of $\kappa$ that this correction is negative for $\Delta<-\sqrt{1/12}$. For $\Delta\lesssim -0.8\Gamma$ we find that the curves of the final action $\langle I\rangle$ as a function of $A_z$ have indeed a negative derivative and decrease (initially) as the micromotion is increased from $A_z=0$. This result does not hold for smaller detunings, for which the mean action always increases, because of higher order terms and the fact that the condition for the validity of this expansion does not accurately hold. The Taylor expansion of the Lorentzian as in \eq{Eq:LorentzianLin2} is relevant subject to the condition
\be \max{v_z} \approx A_z \ll \left[\Gamma^2 +4\Delta^2\right]/(8k|\Delta|)\label{Eq:kvLin2}, \ee
which defines a curve $A_{\rm linear}(\Delta)$ shown in \fig{fig:Imean30}. For $A_z$ obeying this condition, the effect of the micromotion vanishes at first order in $A_z$, and it is small effect overall. In addition, since the term proportional to $\kappa$ was evaluated based on assuming also $p\ll \dot{\bar{z}}$, its relevance requires also that
\be {I} \ll A_z^2 / (12 \nu_z)\label{Eq:IAz2},\ee
(where the numerical factor 12 is approximate), which does not hold for $A_z$ too small. This is why for $-0.8\lesssim \Delta/\Gamma<0$, \eq{Eq:kvLin2} and \eq{Eq:IAz2} cannot be satisfied together, and indeed we find that $\langle I\rangle$ monotonously increases as a function of $A_z$ for these detunings (instead of decreasing initially).

\section{Excess micromotion Expanded in terms of Bessel functions}\label{Sec:Bessel}

To connect our Floquet approach with the excess micromotion treatment of \cite{devoe1989role,Blumel1989, berkeland1998minimization}, starting from the OBE in the form of \eqss{Eq:OBE1}{Eq:OBE2}, we can make a different substitution than that  of \eq{Eq:OBEsubs1}, 
\be\sigma_{ge}'= e^{-i\omega_{\rm L}t}\sigma_{ge},\ee
obtaining, to first order in $\Omega_{\rm R}$, the OBE in the form
\begin{align}
&\dot{\sigma}_{ee} =-\Gamma{\sigma}_{ee} -i\frac{\Omega_{\rm R}}{2} \left(e^{-i\phi}{\sigma}_{ge}' - e^{i\phi}{\sigma}_{eg}'\right),\label{Eq:OBE1c}\\
&\dot{\sigma}_{ge}' = \left(i\Delta-\frac{\Gamma}{2}\right){\sigma}_{ge}' +i\frac{\Omega_{\rm R}}{2} e^{i\phi}.\label{Eq:OBE2d}
\end{align}
Writing the phase in the following form (setting $\phi_{\rm L}=0$),
\be \phi=-\vec{k}\cdot\vec{r}(t)=\beta\cos(2t),\label{Eq:phi2}\ee 
amounts to an expansion of the excess micromotion but not the perturbations about it. We have
\be e^{i\phi}=\sum_n i^n J_n(\beta)e^{i2nt}.\label{Eq:expiphi2}\ee 
Expanding again
\be {\sigma}_{ge}'=\sum_{n}G_{2n}e^{i2n t},\ee
we have
\bem i\sum_{n}2n G_{2n} e^{i2n t}=\left(i\Delta-\frac{\Gamma}{2}\right)\sum_{n}G_{2n}e^{i2n t}\\+i\frac{\Omega_{\rm R}}{2}\sum_n i^n J_n(\beta)e^{i2nt} ,\end{multline}
which implies
\bem G_{2n}=\frac{i^{n+1}(\Omega_{\rm R}/2)J_n(\beta)}{i(2n-\Delta)+\Gamma/2} \\=\frac{i^{n+1}(\Omega_{\rm R}/2)J_n(\beta)\left[\Gamma/2-i(2n-\Delta)\right]}{(2n-\Delta)^2+(\Gamma/2)^2}.\end{multline}
We have
\bem e^{-i\phi}{\sigma}_{ge}' =\sum_{m,n}  i^{-m} J_{m}(\beta) e^{-i2m t} G_{2n}e^{i2n t} \\=\sum_{m,p} i^{-m} J_{m}(\beta) G_{2(m+p)} e^{i2p t},\label{Eq:expiphige}\end{multline}
and substitution in \eq{Eq:OBE1c} gives
\be \dot{\sigma}_{ee} =-\Gamma{\sigma}_{ee} +\Omega_{\rm R}{\rm Im} \left(\sum_{m,p} i^{-m} J_{m}(\beta) G_{2(m+p)} e^{i2p t} \right). \label{Eq:dotsigma_ee2}\ee
Expanding,
\be \sigma_{ee}
={\rm Im}\sum_{n}\rho_{2n} e^{i2n t},
\label{Eq:sigma_eeFloquet2}\ee
 we have
\be \dot \sigma_{ee}
={\rm Im}\sum_{n}i2n\rho_{2n} e^{i2n t},
\label{Eq:dotsigma_eeFloquet2}\ee
and substitution in \eq{Eq:dotsigma_ee2} gives
\be i2n\rho_{2n} =-\Gamma \rho_{2n} +\Omega_{\rm R}\sum_{m} i^{-m} J_{m}(\beta) G_{2(m+n)}, \label{Eq:sigma_eeRecursion}\ee
which can be solved for all $n$.
For the stationary component $n=0$, we get
\be \rho_{0} = \frac{\Omega_{\rm R}}{\Gamma}\sum_{m} i^m J_{m}(\beta) G_{2m}, \label{Eq:rho0}\ee
and the known result for the averaged absorption probability over a micromotion-period (as mentioned in \cite{devoe1989role}), 
\be \rho_{0}= \frac{\Omega_{\rm R}^2}{4}\sum_m \frac{J_m(\beta)^2}{(2m-\Delta)^2+(\Gamma/2)^2}.
\ee

\bibliographystyle{hunsrt}
\bibliography{rf_cooling}

\end{document}